\documentclass[10pt,
  prl,
  twocolumn,
  aps,
  amssymb,
  longbibliography,
  showpacs,
  superscriptaddress,
  floatfix,
  colorlinks,
  linkcolor=blue,
  citecolor=blue,
  anchorcolor=blue
]{revtex4-2}
\usepackage{graphicx}
\usepackage{float}       
\usepackage{dcolumn}
\usepackage{siunitx}
\usepackage{hyperref}
\usepackage{mathtools}
\usepackage{mathbbol}
\usepackage{amsmath}
\usepackage{physics}
\usepackage[caption=false]{subfig}
\usepackage{booktabs}
\usepackage{color}
\usepackage{xcolor}
\usepackage{balance}
\usepackage{braket}

\begin{document}

    \title{Optimizing the dynamical preparation of quantum spin lakes on the ruby lattice}
    
	\author{DinhDuy Vu}\thanks{These authors contributed equally to this work.}
	\affiliation{Department of Physics, Harvard University, 17 Oxford St., Cambridge, MA 02138, USA}
    
	\affiliation{Harvard Quantum Initiative, 60 Oxford St., Cambridge, MA 02138, USA}
    
	\author{Dominik S. Kufel}\thanks{These authors contributed equally to this work.}
	\affiliation{Department of Physics, Harvard University, 17 Oxford St., Cambridge, MA 02138, USA}
    
	\affiliation{Harvard Quantum Initiative, 60 Oxford St., Cambridge, MA 02138, USA}
	\author{Jack Kemp}
	\affiliation{Department of Physics, Harvard University, 17 Oxford St., Cambridge, MA 02138, USA}
	\affiliation{Harvard Quantum Initiative, 60 Oxford St., Cambridge, MA 02138, USA}
	\affiliation{Cavendish Laboratory, University of Cambridge, Cambridge CB3 0HE, United Kingdom}
	\author{Lode Pollet}
	\affiliation{Department of Physics and Arnold Sommerfeld Center for Theoretical Physics (ASC), Ludwig-Maximilians-Universität München, Theresienstrasse 37, München D-80333, Germany}
	\affiliation{Munich Center for Quantum Science and Technology (MCQST), Schellingstrasse 4, D-80799 München, Germany}
	\author{Chris R. Laumann}
    \affiliation{Department of Physics, Harvard University,  17 Oxford St., Cambridge, MA 02138, USA}
	\affiliation{Department of Physics, Boston University, 590 Commonwealth Avenue, Boston, Massachusetts 02215, USA}
    \affiliation{Max-Planck-Institut f\"{u}r Physik komplexer Systeme, 01187 Dresden, Germany}
    
	\author{Norman Y. Yao}
	\affiliation{Department of Physics, Harvard University, 17 Oxford St., Cambridge, MA 02138, USA}
	\affiliation{Harvard Quantum Initiative, 60 Oxford St., Cambridge, MA 02138, USA}

	\begin{abstract}
        Quantum spin liquids are elusive long-range entangled states.
        Motivated by experiments in Rydberg quantum simulators, recent excitement has centered on the possibility of dynamically preparing a state with quantum spin liquid correlations even when the ground-state phase diagram does not exhibit such a topological phase. 
        Understanding the microscopic nature of such quantum spin ``lake'' states and their relationship to equilibrium spin liquid order remains an essential question. 
Here, we extend the use of approximately symmetric neural quantum states for real-time evolution and directly simulate the dynamical preparation in systems of up to $N=384$ atoms. 
We analyze a variety of spin liquid diagnostics as a function of the preparation protocol and optimize the extent of the quantum spin lake thus obtained.
In the optimal case, the prepared state shows spin-liquid properties extending over half the system size, with a topological entanglement entropy plateauing close to $\gamma = \ln 2$. We extract two physical length scales $\lambda_e$ and $\xi_m$ which constrain the extent of the quantum spin lake $\ell$ from above and below.
	\end{abstract}
	
	\maketitle

Quantum spin liquids (QSLs) are exotic phases of matter with macroscopic entanglement, fractionalized excitations, and emergent gauge symmetry~\cite{anderson1973resonating, kitaev2006tee, savary2016quantum}. 
Realizing and characterizing spin-liquid ground states remains a long-standing challenge across quantum materials and quantum simulation~\cite{shimizu2003qsl, wen2019kitaev_qsl, verresen2021prediction, semeghini2021probing, Knolle2019review}. 
Although many material candidates have been identified, positive signatures of spin-liquid behavior remain scarce~\cite{Czajka2021Oscill, Zeng2024Dirac, Zheng2025Kagome, Bruin2022ThermalHall, Czajka2023Planar}.

\begin{figure}[ht!]
    \centering
    \includegraphics[width=0.95\linewidth]{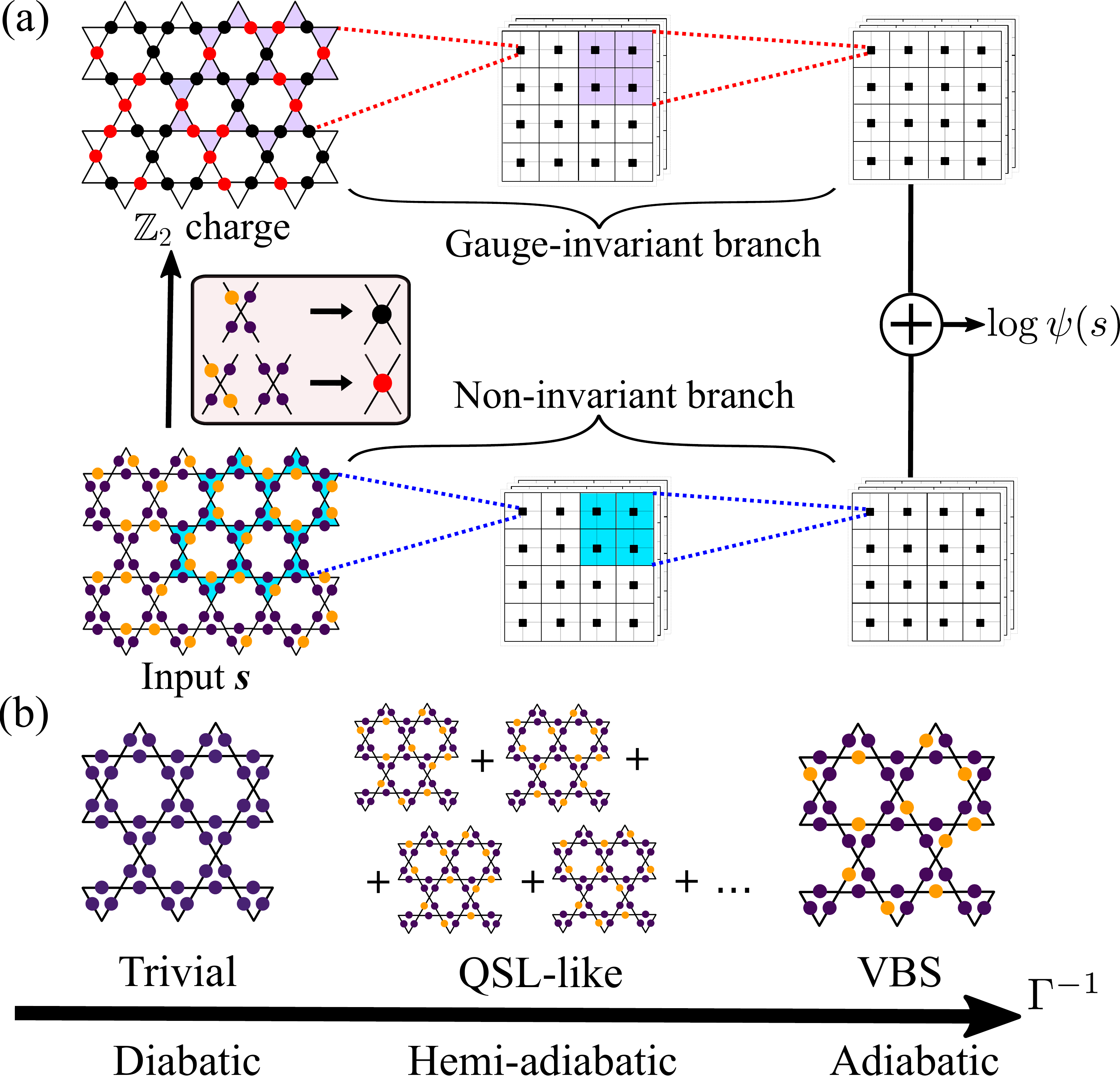}
    \caption{(a) Approximately gauge-symmetric neural network architecture, consisting of a gauge-symmetric branch (upper) and a non-symmetric branch (lower). The Rydberg configuration (purple/orange for ground/Rydberg states) enters the non-symmetric branch, while black/red dots in the gauge-symmetric branch denote the $-1/+1$ values of the parity of Rydberg excitations around a vertex.
    (b) Dynamical preparation of a QSL by ramping the Hamiltonian at rate $\Gamma$ to excite atoms from the ground state to the Rydberg state subject to Rydberg blockade.} 
    \label{fig:cover}
\end{figure}

Quantum simulators provide alternative routes to spin liquids. 
Digital platforms can prepare spin-liquid states using projective measurements and feedback~\cite{googleTC, iqbal2024non,iqbal2025qutrit}, while analog simulators typically rely on coherent adiabatic preparation~\cite{farhi2000quantum,hamma2008adiabatic,islam2013emergence,barbarino2020preparing,ebadi2021quantum,mambrini2024quantum,cookmeyer2024qsl, bintz2024dirac}. 
In the latter approach, the Hamiltonian is slowly tuned so that the system follows its instantaneous ground state into the spin-liquid phase. 
Recent Rydberg-simulator observations suggest an even more surprising possibility~\cite{semeghini2021probing}: one may dynamically ramp into a spin-liquid-like state even when the underlying Hamiltonian \emph{does not} host a QSL ground state~\cite{verresen2021prediction}.

This possibility has sparked intense debate about the nature of such dynamically generated states~\cite{Giuliano2022dynamic,sahay2023lakes, mauron2025predicting,patil2025qmc,Wang2025classical, gjonbalaj2025shortcut}. 
Physical arguments~\cite{sahay2023lakes, gjonbalaj2025shortcut} suggest that spin-liquid correlations can emerge over finite length scales up to a cutoff $\ell$, defining a ``quantum spin lake.'' 
By contrast, equilibrium quantum Monte Carlo has questioned the state's quantum coherence and found strong area-law scaling of the Wilson loop~\cite{Wang2025classical}, while time-dependent variational Monte Carlo reported a non-quantized topological entanglement entropy~\cite{mauron2025predicting}.

In this Letter, we reconcile these perspectives using large-scale neural-quantum-state (NQS) simulations of dynamical spin-liquid preparation. 
Our main results are three-fold. 
First, using the approximately symmetric neural network architectures~\cite{kufel2025} [Fig.~\ref{fig:cover}(a)], we perform the highest-fidelity simulations of the experimental ramp protocol to date, reaching larger systems ($N\le384$), longer times, and more general ramp protocols. 
Second, we analyze spin-liquid diagnostics versus ramp rate: slow ramps revert to the adiabatic strategy and reach the actual non-spin-liquid ground state, while fast ramps remain near the trivial initial state. 
At intermediate rates, we find quantum spin lakes [Fig.~\ref{fig:cover}(b)] whose topological entanglement entropy matches that of a $\mathbb{Z}_2$ QSL. Finally, using an anyon-gas framework, we quantify the maximum spin-lake extent as well as the optimal preparation protocol. In a broader context, our work demonstrates that topological correlations can be dynamically generated over mesoscopic scales even when the physical Hamiltonian does not host the corresponding topological phase. This applies to other exotic states such as those in  recent $U(1)$-spin liquid experiments \cite{geim2026, karch2026}.

\emph{Rydberg Hamiltonian and neural quantum states.}---
Consider a Rydberg quantum simulator with  $N$ neutral atoms arranged on a ruby lattice (i.e. links of a kagome lattice) [Fig.~\ref{fig:cover}].
Each atom behaves as an effective two-level system defined by the electronic ground state and a highly excited Rydberg state: $\{ \ket{g},\ket{r} \}$.
In the presence of an optical drive addressing the ground-Rydberg transition, the many-body Hamiltonian governing the system is:
	\begin{equation}
		H = -\frac{\Omega}{2}\sum_i (b_i + b_i^\dagger) + \sum_{ij} V_{ij}n_i n_j - \delta \sum_i n_i 
        \label{eq:rydberg}
	\end{equation}
where $b_i^\dagger$ creates a Rydberg excitation at ruby-lattice site $i$, $n_i = b_i^\dagger b_i$, $\Omega$ is the Rabi frequency of the optical drive with detuning $\delta$, and $V_{ij}=\Omega (R_b/R_{ij})^6$ where $R_b$ is the blockade radius. 

The Hamiltonian \eqref{eq:rydberg} is best understood by starting from the idealized `PXP' limit, which corresponds to an infinite repulsion within the blockade radius $R_b=2a$ ($a$ denotes the spacing between atoms within a triangle) and no further interaction tail~\cite{verresen2021prediction}. 
In this constrained space, no more than one of the four atoms surrounding each kagome vertex can occupy the Rydberg state.
The PXP system exhibits $\mathbb{Z}_2$ spin-liquid ground states for $1.4\lesssim \delta/\Omega \lesssim 2.0$, which may heuristically be understood as superpositions of all the states satisfying the Gauss law-like constraint that each kagome vertex is surrounded by one Rydberg.
More precisely, the PXP spin liquid has been shown~\cite{verresen2022unify} to be adiabatically connected to the ground state of the toric-code-like stabilizer model, 
\begin{equation}
    \raisebox{-30pt}{\includegraphics[width=0.6\linewidth]{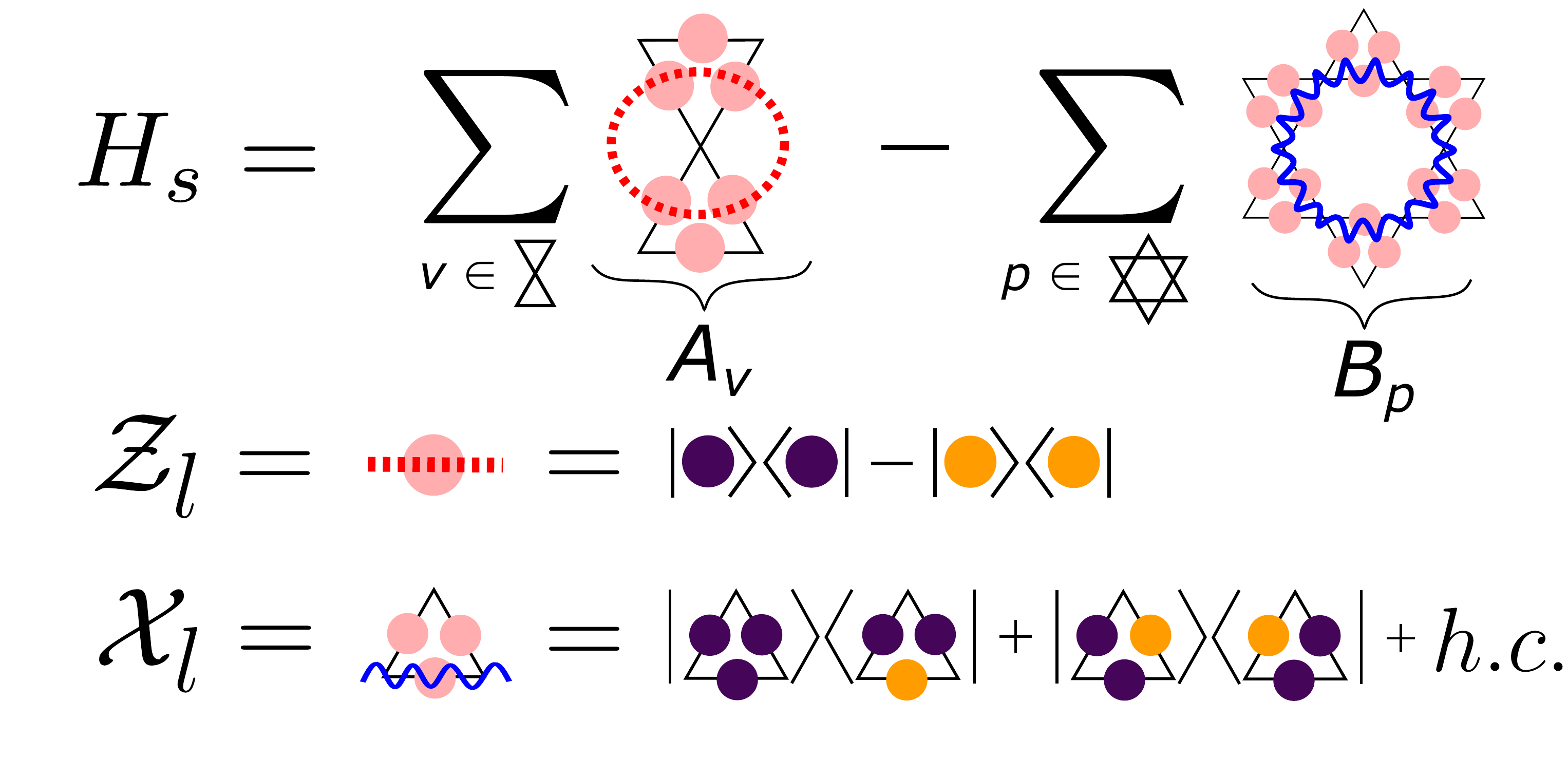}}.
    \label{eq:stabilizerlimit}
\end{equation}
Within this description, one restricts the Hilbert space of three atoms on a triangle to the single-excitation subspace $\{\ket{ggg},\ket{rgg},\ket{grg},\ket{ggr}\}$, consistent with strong intra-triangle blockade. The dashed red line corresponds to a $\mathcal{Z}$ operator that assigns $+1$ to the ground state (purple dot) and $-1$ to the Rydberg-state (orange dot), while the wavy blue line corresponds to an $\mathcal{X}$ operator that creates, annihilates, or shuffles a ``Rydberg dimer'' around the triangle~\cite{verresen2021prediction,verresen2022unify}. 
The two terms in the stabilizer Hamiltonian correspond to: (i) the sum of $A_v$--product of $\mathcal{Z}$ around one vertex and (ii) the sum of $B_p$--product of $\mathcal{X}$ over one hexagonal plaquette.
Physically, the former enforces the local Gauss' law, while the latter induces resonances among  Gauss’-law-obeying configurations. We note that the spin liquid phase on the Rydberg ruby lattice can also be connected to the ground state of another stabilizer Hamiltonian with a different definition of the $B_p$ terms \cite{Tarabunga2022}. However, as shown latter, our variational ansatz only explicitly uses the diagonal $A_v$ terms, so this variation of stabilizer wavefunctions does not alter our ansatz design.

In order to study the broad class of $\mathbb{Z}_2$ QSLs on ruby Rydberg lattices connected to the ground state of Eqn.~\eqref{eq:stabilizerlimit}, we utilize neural quantum states (NQS) tailored to the approximate gauge group symmetry \cite{kufel2025}. Within the NQS framework, a quantum state is represented as $\ket{\psi}=\sum_{s}\psi(s)\ket{s}$, 
where a neural network maps a basis configuration $s$ to a complex amplitude $\psi(s)$~\cite{carleo2017solving}. The past several years have witnessed rapid progress in NQS methods,  with state-of-the-art results obtained for quantum ground states \cite{carleo2017solving, choo2018symmetries, luo2019backflow, choo2019two, roth2023high}, steady states of open systems \cite{Torlai2018latent, carrasquilla2019reconstructing, Nomura2021purify, Wagner2024thermo}, and dynamics problems~\cite{Schmitt2020dynamics, mendessantos2023resolve, Schmitt2022phase}.
In the context of Rydberg quantum simulation, a key advantage of NQS-based methods compared to other numerical approaches is the ability to directly simulate experimentally relevant geometries and system sizes~\cite{mauron2025predicting}.

Since all $A_v$ and $B_p$ commute, $H_s$ exhibits an exponentially large group of local symmetries generated by the $B_p$, i.e. $B_p\ket{\psi_\text{stabilizer}} = \ket{\psi_\text{stabilizer}} \forall p$, making it naturally suited for a gauge-symmetric neural network architecture~\cite{luo2021gauge,luo2023gauge}. 
In Fig.~\ref{fig:cover}(a), the symmetric branch takes $A_v$ values as input and accordingly produces $\ket{\psi_\text{sym}}$ such that $B_p\ket{\psi_\text{sym}}=\ket{\psi_\text{sym}}$.
However, for a generic state in the same phase as the stabilizer state, the $B_p$ are no longer exact symmetries.
We thus use an  approximately symmetric architecture which includes a parallel branch with no symmetry restrictions~\cite{kufel2025}. 
In practice, both the gauge-symmetric and non-symmetric branches are implemented as three-layer convolutional networks with 8 features per hidden layer, polynomial activations, and a final average-pooling step~\cite{suppinfo}.  
We further stabilize the network by incorporating a 3-parameter translationally invariant branch (not shown) which directly parametrizes the probabilities of violating Rydberg blockade (see End Matter and \cite{suppinfo}).
The outputs from the three branches are summed to produce the logarithmic amplitude of each basis state $\log \psi_s$.
    
\begin{figure}
    \centering
    \includegraphics[width=0.49\textwidth]{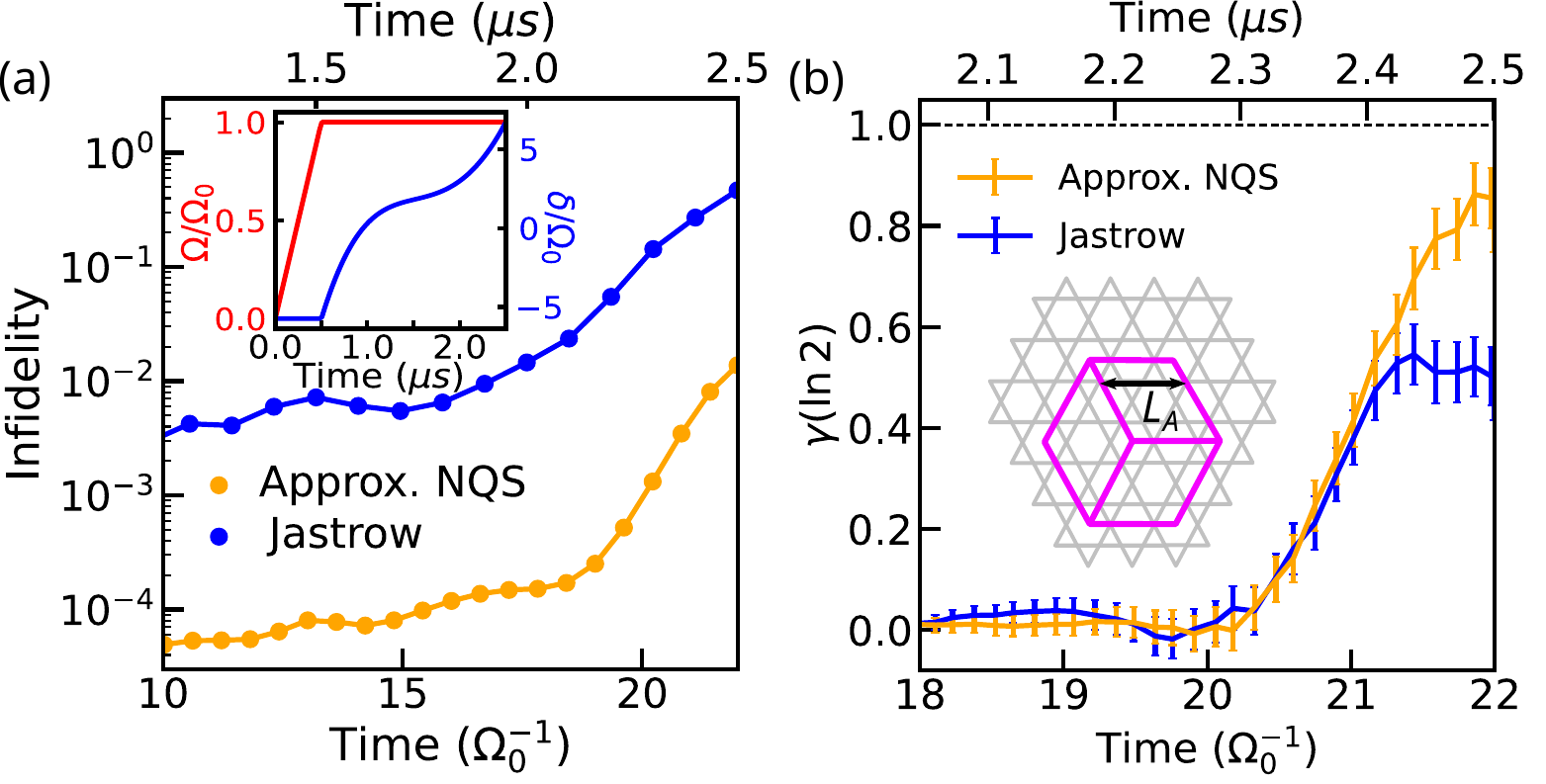}
    \caption{
    (a) $N=24$ with periodic boundary conditions: Infidelity of the approximately-symmetric NQS (orange dots) and Jastrow ansatz in Ref.~\cite{mauron2025predicting} (blue squares) as compared with exact diagonalization. The ramping profile is taken from Ref.~\cite{semeghini2021probing} (inset) with $\Omega = \Omega_0$ and $\delta \approx 6.7~\Omega_0$ at the end of the ramp.
    (b) $N=219$ with open boundary conditions: Comparison of the TEE ($\gamma$) from approximately-symmetric NQS (orange) and Jastrow ansatz (blue) over the subsystem (magenta borders) of scale $L_A$ shown in the inset. The values of $\gamma$ at the end of the ramp are $\approx 0.38$ for Jastrow and $\approx 0.60$ for approximately-symmetric NQS, compared with the expected value $\approx 0.69$ ($\ln 2$) indicated by the dashed line.}
    \label{fig:dynamic_benchmark}
\end{figure}

\textit{Preparing a dynamical quantum spin liquid}---Unfortunately, the PXP spin-liquid ground state appears to be unstable to the reintroduction of the long-range tails of the van der Waals interaction (i.e.~$V_{ij} \sim  1/R_{ij}^6$). 
This was shown by tensor network methods studying the zero-temperature phase diagram of Eqn.~(\ref{eq:rydberg})~\cite{verresen2021prediction}, which we corroborate using large-scale ground-state NQS simulations in the End Matter. 
Intriguingly, signatures of spin-liquid correlations were observed in the experimental dynamical protocol starting from the product state $\ket{g}^{\otimes N}$ and ramping both $\Omega(t)$ and $\delta(t)$ [inset, Fig.~\ref{fig:dynamic_benchmark}(a)]. In order to simulate the dynamical preparation protocol, we utilize our approximately-symmetric NQS within the time-dependent variational principle (TDVP)~\cite{saraceno1981tdvp, suppinfo}.
%
As a benchmark, we begin by directly simulating the ramping profile implemented in the Rydberg experiment [inset, Fig.~\ref{fig:dynamic_benchmark}(a)]~\cite{semeghini2021probing}; we note that a recent work has also explored these dynamics via TDVP using a decorated Jastrow ansatz~\cite{mauron2025predicting}. 
Working at small system sizes ($N=24$), Figure~\ref{fig:dynamic_benchmark}(a) depicts the state infidelity as a function of the ramp time for both ansatzes compared to exact diagonalization. 
The approximately-symmetric NQS exhibits an infidelity nearly two orders of magnitude lower across the entirety of the ramping protocol; this is especially important near the end of the dynamical spin liquid preparation where the state obtained by the Jastrow ansatz exhibits a $\sim 50\%$ infidelity, while the state obtained via NQS exhibits a $\sim 1\%$ infidelity~\footnote{We note that directly optimizing with respect to the exact state yields an additional order-of-magnitude improvement in the infidelity for the approximately-symmetric NQS. This improvement is not seen for the Jastrow ansatz~\cite{suppinfo}.}.

Building on these fidelities, we now directly simulate the preparation dynamics on significantly larger system sizes ($N=219$) using the precise experimental geometry (inset, Fig.~\ref{fig:dynamic_benchmark}(b)). To quantify the topological nature of the prepared state, we track the topological entanglement entropy (TEE) of the state.
Even though a protecting energy gap is not present in this context, a nearly quantized TEE still distinguishes good spin liquid-like properties from poor ones on mesoscopic length scales.
For this geometry, Jastrow ansatz simulations observe a gradual rise in the TEE at scale $L_A=4$ (about half the system size) as a function of the ramping time [blue curve, Fig.~\ref{fig:dynamic_benchmark}(b)]; however, the TEE does not saturate anywhere near the value expected ($\gamma = \ln 2$)  for a  $\mathbb{Z}_2$ spin liquid, leading previous authors to the conclusion that the state does not exhibit a true $\mathbb{Z}_2$ topological order~\cite{mauron2025predicting}.  
For early and intermediate times the
NQS and Jastrow results track each other closely; they begin to separate around the time when the Jastrow TEE starts to saturate, and this difference becomes most pronounced near the end of the ramp, where the NQS continues rising toward the $\ln 2$  value expected for a $\mathbb{Z}_2$ spin liquid [orange curve, Fig.~\ref{fig:dynamic_benchmark}(b)]. 

\begin{figure}
    \centering
    \includegraphics[width=1.02\linewidth]{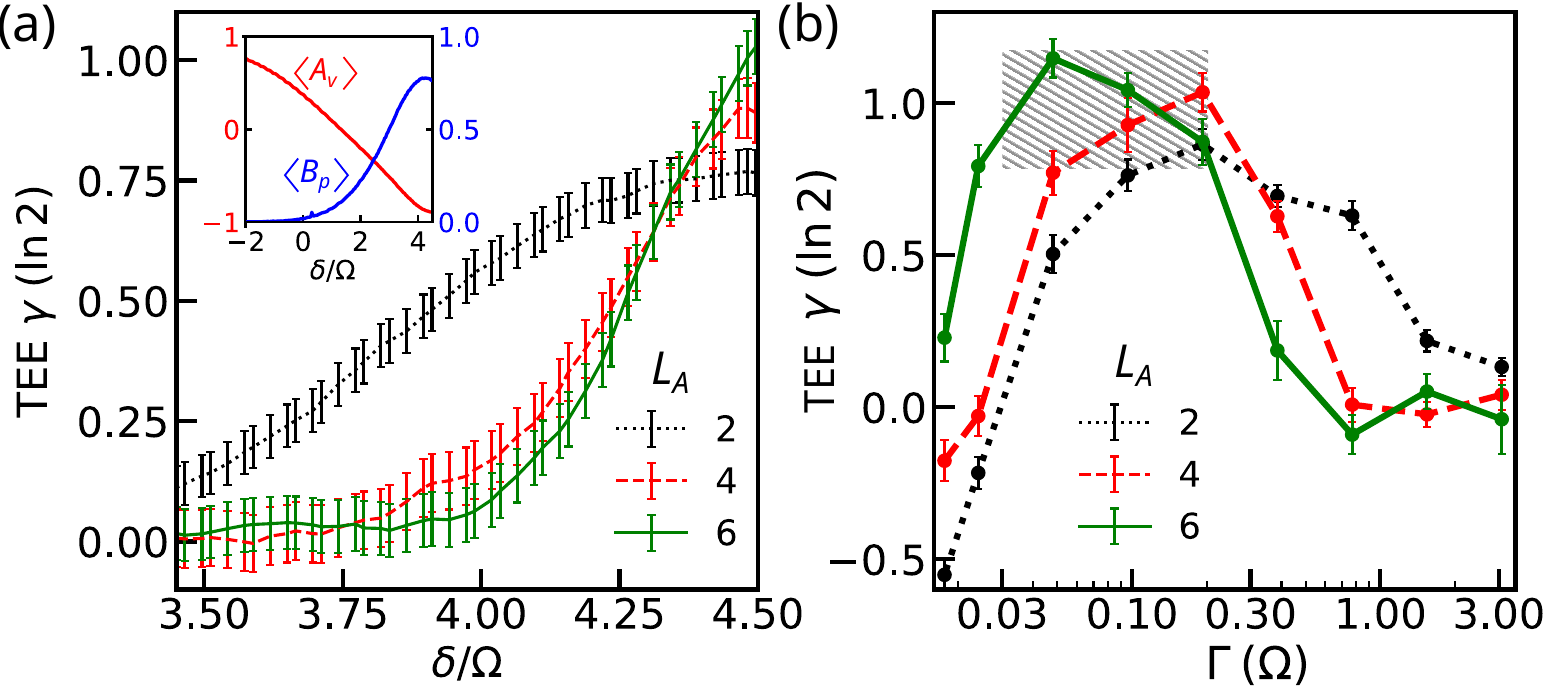}
    \caption{Optimizing the preparation protocol for a spin lake. (a) Topological entanglement entropy at ramping rate $\Gamma\approx 0.096~\Omega$ while varying the final $\delta$. The inset shows the evolution of stabilizers $A_v$ and $B_p$ with final $\delta$ over a wider range. (b) Topological entanglement entropy over different subsystems (see Fig.~\ref{fig:partition}(b) in \cite{suppinfo}) at fixed final $\delta/\Omega = 4.5$ while varying the ramp rate $\Gamma$. The hatched region indicates a proxy for finite-size spin-liquid-like behavior over the range $\Gamma \sim 0.03$ to $\sim 0.2$ with a TEE close to $\ln 2$. The subsystem size $L_A$ is depicted in the inset of Fig.~\ref{fig:dynamic_benchmark}(b).}
    \label{fig:drawing_3}
\end{figure}

\textit{Optimizing the preparation protocol}---
We take the observation of TEE $\gamma \approx \ln 2$ as partial evidence that the state prepared by the dynamical protocol is indeed a `spin lake' state~\cite{sahay2023lakes} for $\ell \gtrsim L_A = 4$.  
We now optimize the preparation protocol in order to maximize the spin-lake size $\ell$. 
To simplify the investigation, we use linear ramps and a larger Rydberg-atom lattice with periodic boundary conditions and $N=384$ atoms. The linear ramp is engineered so that the Rabi frequency $\Omega$ and blockade radius $R_b=2.4~a$ are fixed while the detuning $\delta(t)=\Omega(-2+\Gamma t)$ is swept from the trivial phase $\delta/\Omega = -2$ at a constant rate $\Gamma$.
In the inset of Fig.~\ref{fig:drawing_3}(a), we roughly tune the protocol by fixing the rate $\Gamma \approx 0.1~\Omega$ and measuring $\langle A_v\rangle$ and $\langle B_p \rangle$ as a function of the final detuning $\delta$ of the ramp. We find that $\langle A_v \rangle$ approaches $-1$ and $\langle B_p \rangle$ approaches $+1$, as expected of the stabilizer Hamiltonian ground state spin liquid, around $\delta/\Omega \approx 4.5$, squarely in the ground state valence bond solid (VBS) phase. 
We confirm the topological nature of the prepared state by computing the scaling of TEE with the subsystem size $L_A$ in Fig.~\ref{fig:drawing_3}(a) (also see End Matter for the Fredenhagen-Marcu order parameter). Because the dynamically
prepared state is expected to exhibit spin-liquid-like correlations only up to a finite length scale independent of the system size, subsystem-scaling diagnostics are the natural tool for determining over what spatial extent the state behaves topologically.
Near the optimal $\delta/\Omega \approx 4.5$, we observe a consistent improvement of TEE with larger subsystems up to $L_A = 6$ (the largest size available without touching the boundaries). 
This suggests that the spin lake prepared by sweeping into the valence bond solid (VBS) phase is much larger than the lattice scale and comparable with the lattice size ($L=16$ kagome links). Remarkably, as demonstrated in Fig.~\ref{fig:drawing_3}(b), the extent of the spin lake is also governed by the ramp rate. Fixing the final detuning at $\delta/\Omega = 4.5$, the TEE first increases as the ramp rate is slowed, reaching a value near $\ln 2$, before subsequently decreasing back to zero for adiabatic ramps. Focusing on the largest subsystem $L_A=6$, one can identify an optimal ramping rate around which the TEE is close to $\ln 2$ [the hatched region in Fig.~\ref{fig:drawing_3}(b)]. 

\textit{Characterizing the prepared spin lake.}---The TEE provides a numerically concise, basis-independent diagnostic of spin-liquid correlations.
Physically, however, spin lakes should be understood in terms of correlated gases of $e$- and $m$-anyons, which we define through violations of the $A_v$ and $B_p$ operators referenced to the stabilizer Hamiltonian~\eqref{eq:stabilizerlimit}. 
Each anyon gas is characterized by a mean spacing, $\lambda_e$ or $\lambda_m$, and a pair-correlation length, $\xi_e$ or $\xi_m$, which governs the distance over which locally produced or annihilated pairs have separated by the time of measurement. 
These length scales can be extracted by measuring equal-time Wilson loops and strings at any given time during the ramp \cite{Gregor2011, xu2025critical}.

The initial state of the ramp is the product $\ket{g}^{\otimes N}$, which has an $e$ at every kagome vertex ($A_v = +1$). 
These are projected out by the ramp -- in the adiabatic limit, there are only weak virtual pair-fluctuations of $e$'s in the final VBS state.
For a slow but non-adiabatic ramp, the $e$'s are annihilated in local pairs and the residual $e$'s then diffuse apart and annihilate if they find each other.
We thus expect a dilute gas of residual $e$-anyons such that $\xi_e > \lambda_e$.

\begin{figure}
		\centering
		\includegraphics[width=\linewidth]{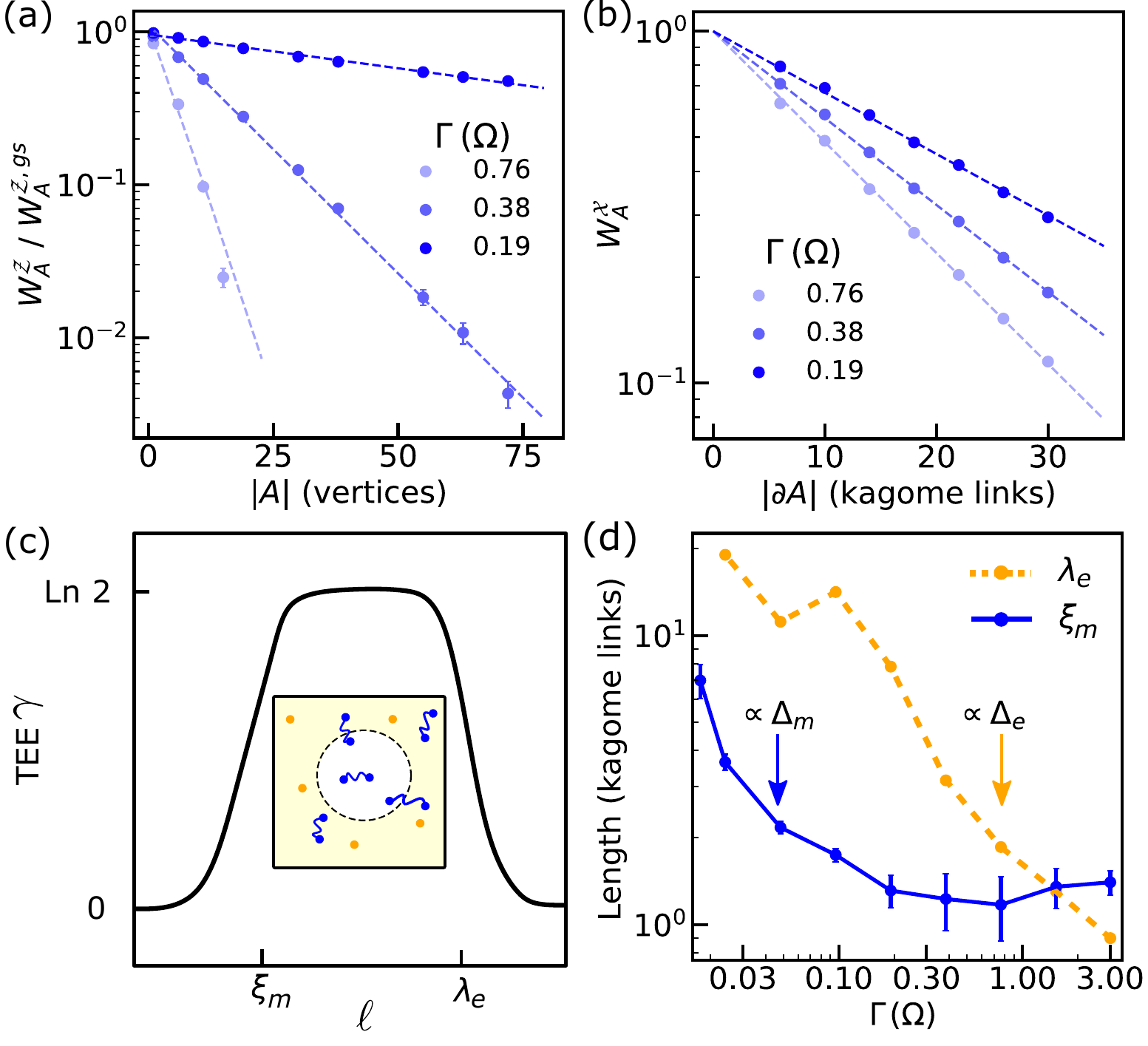}
			\caption{(a) The area-law scaling of $W_A^{\mathcal{Z}}/ W_A^{\mathcal{Z}, gs}$ at different ramp rates. (b) The perimeter-law scaling of $W_A^\mathcal{X}$. (c) Pictorial depiction of TEE with respect to the subsystem length scale $\ell$. TEE approaches $\ln 2$ for $\lambda_e > \ell > \xi_m$. The inset shows the free $e$-gas (orange circles) with average distance $\lambda_e$ and tightly bound $m$-pairs
        (connected blue dot pairs) with pair length $\xi_m$. (d) The spatial upper bound $\lambda_e$ and lower bound $\xi_m$ of the spin lake as functions of ramp rate. The adiabatic limit where the length scale starts to grow significantly is set by the energy scales $\Delta_e$ and $\Delta_m$ of the respective excitation species.}
		\label{fig:drawing_4}
	\end{figure}

We can check the consistency of this picture from several equal-time observables.
The residual defect spacing $\lambda_e$ can be estimated by $1/\lambda_e^2 = 2(n_e - n_e^{gs})$, where $n_e = \left(1+\langle A_v \rangle\right)/2$. 
Assuming the residual $e$'s form a nearly-free gas, we should also find an area-law decay of the $\mathcal{Z}$-Wilson loops, since
\begin{equation}\label{eq:area-decay}
    \braket{W_A^{\mathcal{Z}}}
\approx
\braket{\prod_{v\in A} A_v}
\approx
\prod_{v\in A}\braket{A_v}
=(-1)^{|A|} e^{-|A|/\lambda_e^2}.
\end{equation}
As shown in Fig.~\ref{fig:drawing_4}(a), after dividing out the corresponding loops evaluated in the ground state of the Hamiltonian at the end of the ramp, $W_A^{\mathcal{Z}}/W_A^{\mathcal{Z},gs}$ exhibits pure area-law scaling. 
The $\lambda_e$ extracted from these two approaches are identical to within numerical error~\cite{suppinfo}. 

In principle, the area-law should revert to a perimeter law beyond the pair-correlation length $\xi_e$ -- it turns out that for all near-optimal ramps, this $\xi_e$ is larger than the accessible system sizes.
We measure $\xi_e$ using the $\mathcal{X}$-string operators, $\langle S_L^\mathcal{X}\rangle \propto e^{-L/\xi_e}$.
Numerically, for the optimal ramp rate $\Gamma\approx 0.096~\Omega$, we find $\xi_e > 20$~kagome links while the largest system size is $L = 16$. 
For comparison, $\lambda_e \approx 10$ for the optimal ramp. Even considering an infinite system, at the length scale $\xi_e$ where the area-to-perimeter crossover happens, the Wilson loop has dropped to such a low value that it is numerically nearly impossible to observe the scaling law.

We now consider the $m$-anyons. 
The essential claim of the spin-lake picture is that the timescale for producing $m$-anyon pairs is much longer than a hemi-adiabatic ramping time; consequently, relatively few $m$-anyon pairs are produced, with a short correlation length at the end of the ramp. Numerically, at the optimal ramp rate, we find $\lambda_m \approx 3.0$ from $1/\lambda_m^2 = 2n_m$ and $n_m = \left(1-\langle B_p \rangle\right)/2$, and $\xi_m \approx 1.7$ from the decay of open $\mathcal{Z}$-strings. 
Since $\xi_m$ is much shorter than the loops we measure, the $\mathcal{X}$-Wilson loops should exhibit perimeter-law decay, as supported by Fig.~\ref{fig:drawing_4}(b).

In summary, we find the hierarchy
$
\xi_e > \lambda_e > \lambda_m > \xi_m .
$
We expect to find the TEE approaching the spin-liquid value only for a region of length scale $\ell$, where $\lambda_e \gtrsim \ell \gtrsim \xi_m$.
$\ell$ must be larger than $\xi_m$ so that the contribution of $m-$anyon pairs arises only from crossings at the boundary. This is not possible for very slow ramps, when a VBS is formed.
On the other side, $\ell$ must be smaller than $\lambda_e$ to ensure that the number of enclosed unpaired $e-$anyons is negligible, preserving the Gauss’ law character of the state. This is not possible for very fast ramps, when the final state is trivial. 
Putting these two bounds together, only for intermediate ramps can one obtain a separation of scales between $\lambda_e$ and $\xi_m$ and observe topological correlations for spatial regions satisfying $\lambda_e \gtrsim \ell \gtrsim \xi_m$ [Fig.~\ref{fig:drawing_4}(c)].

The existence of such a regime relies on the separation between the energy gaps $\Delta_e$ and $\Delta_m$, which control the nonadiabatic production of $e$- and $m$-anyons during the ramp. Although these gaps are only loosely defined---since the Rydberg Hamiltonian governing the dynamics [Eq.~(\ref{eq:rydberg})] differs from the stabilizer Hamiltonian [Eq.~(\ref{eq:stabilizerlimit})]---we estimate them as the energy cost of open-string operators acting on the optimally prepared state, evaluated with respect to the final Hamiltonian. We note that the separation of these energy scales roughly corresponds to the crossover scales in $\lambda_e$ and $\xi_m$ as functions of the ramp rate $\Gamma$ [Fig.~\ref{fig:drawing_4}(d)]~\footnote{Quantitatively, we find $\Delta_e \approx 3.09~\Omega$ and $\Delta_m \approx 0.19~\Omega$ for the spin lake state prepared at the optimal ramp rate.}, and that the largest separation of length scales is achieved for $\Delta_e \gtrsim \Gamma \gtrsim \Delta_m$. Compared with the TEE measure in Fig.~\ref{fig:drawing_3}(b), Fig.~\ref{fig:drawing_4}(d) qualitatively reproduces the optimal ramp rate at different length scales.

\emph{Outlook.}---Our large-scale NQS-based study clarifies how to optimally generate substantial quantum spin-lake states and provides guidance on next-generation Rydberg simulators with larger lattices and longer coherence times \cite{manetsch2025tweezer}.
The NQS architecture can be generalized to other types of topological orders such as the $\mathbb{Z}_3$ \cite{Giudice2022trimer} and especially $U(1)$ \cite{geim2026} quantum spin liquid where there exists no stable equilibrium analog in 2D. 
Moreover, non-Abelian topological orders \cite{iqbal2024non, xu2024non} can also be captured within our framework, where it would be very interesting to understand how the finite spin lake length scale manifests in real-time anyon braiding protocols.
Finally, our simulation can extend to the thermalization dynamics after preparation, revealing the stability of dynamical spin liquids.
	
\begin{acknowledgements}
	\textit{Acknowledgements}---We gratefully acknowledge the insights of and discussions with Marcus Bintz, Johannes Feldmeier, Brenden Roberts, Rahul Sahay, Giulia Semeghini, and Peter Zoller. This work was supported by the U.S. Department of Energy via the Office of Science,
National Quantum Information Science Research Centers, Quantum Systems Accelerator and by the NSF (PHY2608240). J. K. acknowledges support from NSF through the Center for Ultracold Atoms and via EPSRC Grant No. EP/V062654/1. D. S. K. acknowledges support from a Generation-Q AWS fellowship. L. P. acknowledges support from the Deutsche Forschungsgemeinschaft (DFG, German Research Foundation) under Germany’s Excellence Strategy—EXC-2111—Project No. 390814868 and financial support from DFG No. 530111096-ANR-23-CE30-0018. The project and research is part of the Munich Quantum Valley, which is supported by the Bavarian state government with funds from the Hightech Agenda Bayern
Plus. L. P. thanks Harvard University for its hospitality. C. R. L. acknowledges the Gutzwiller Fellowship at Max Planck Institute for the Physics of Complex Systems and the partial support of SFB 1143 and ct.qmat. N. Y. Y. acknowledges support from the Microsoft Quantum Pioneers Program.

	\end{acknowledgements}

    \bibliography{references.bib}

    \section*{End Matter}
    \twocolumngrid
    \emph{Ground state phase diagram}--- Using NQS we can map out the $T=0$ phase diagram of the full $1/r^6$ Rydberg model. This phase diagram is under active debate and provides a baseline for our later dynamical study. Ref.~\cite{verresen2021prediction}, using iDMRG on infinite cylinders, argued that the full $1/r^6$ model does not host a QSL at any value of detuning at the largest truncation distances (unlike its PXP Hamiltonian limit). However, on finite cylinders the long-range interactions must be cut off along the compact direction, and the putative QSL depends on the cylinder width. We revisit this problem with NQS, which can accommodate arbitrary boundary conditions and avoid truncating long-range interactions. 
    
    Obtaining the ground state of the Rydberg-atom model involves two principal challenges: (i) The initial state is significantly far from the true ground state, owing to the large number of blockade violations pushing up the energy; starting from such a high-energy state can cause the neural network to become unstable in the subsequent variational training. (ii) For $\delta \gg \Omega$, the solid-liquid energy splitting is small, and as a result, a randomly initialized ansatz is prone to becoming trapped in a liquid-like local minimum. 
    
    We resolve both issues by adding a mean-field component into the ansatz given by
    \begin{equation}\label{eq:mean_field}
		\log \psi(s) = \theta^{MF}(s) + \theta^{NN}(s),
	\end{equation}
	where $\theta^{NN}(s)$ is the output of the approximately gauge-symmetric neural network and $\theta^{MF}(s)$ is the mean-field ansatz $\theta^{MF}(s) = \sum_i s_i \left(A + \sum_{j:R_{ij} < R_b } B_{R_{ij}} s_j\right) $  with $R_b$ denoting the blockade radius and $s_i = 1 (-1)$ representing the ground state (Rydberg state) at site $i$. Taking into account equal distances, $\theta_{MF}$ is characterized by three parameters: $A$ and two distinct values for $B_{R_{ij}}$. 
    
    To address issue (i), we first pretrain the mean-field component by fixing all parameters in $\theta^{NN}$ in Eqn.~\eqref{eq:mean_field} and optimizing only those in $\theta^{MF}$. Starting from a random initialization, the $B_{R_{ij}}$ parameters quickly acquire large negative values, effectively capturing the low probability of Gauss-law–violating configurations and yielding an energy close to the ground state. This pretrained state provides a stable foundation for subsequent refinement of the deep neural network.

\begin{figure}
		\centering
		\includegraphics[width=0.35\textwidth]{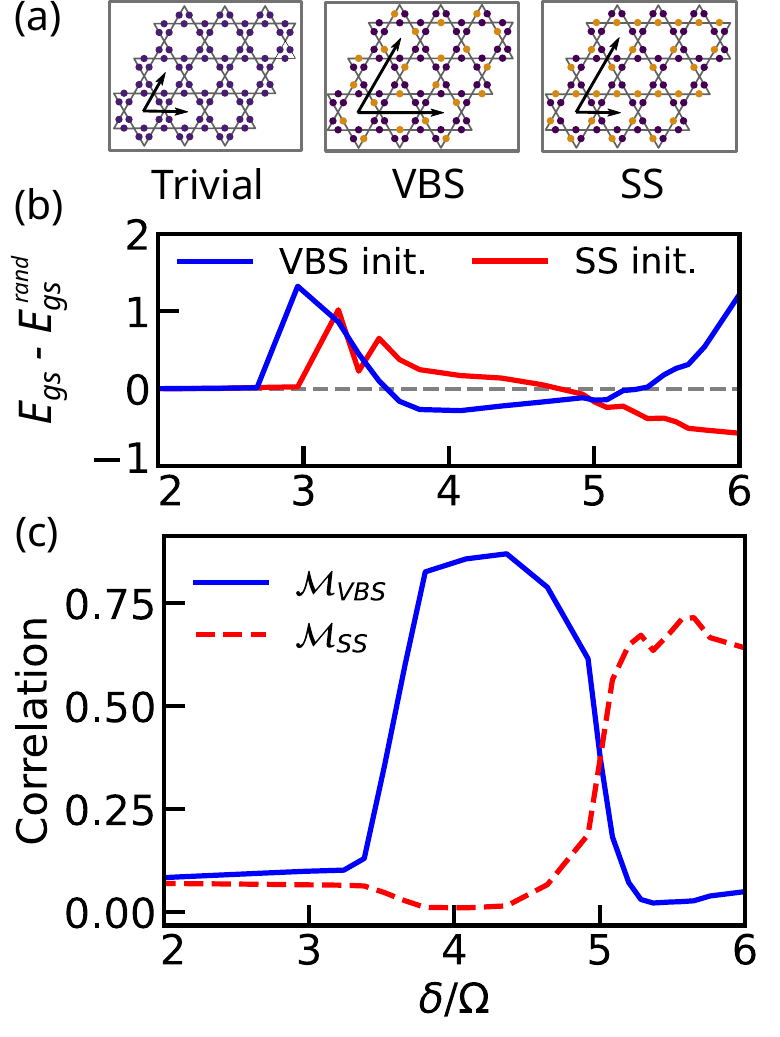}
		\caption{Ground state phase diagram as a function of $\delta/\Omega$. (a) Excitation pattern in the three phases: symmetric trivial, symmetry-breaking valence bond solid (VBS), and symmetry-breaking stripe solid (SS). The unit vectors in each phase are shown with arrows. The VBS state doubles the unit cell along both lattice unit-vector directions, while the SS does so in one direction. (b) Ground state energy with VBS and SS initializations offset by $E_{gs}$ computed with random initialization. (c) Correlations [as defined by Eqn.~\eqref{eq:order_params}] detecting the symmetry-breaking pattern of the VBS (blue solid) and SS (red dashed) phases.}
		\label{fig:GS}
    \end{figure}

    To mitigate issue (ii), we guide the search toward specific symmetry-breaking patterns while still letting the lowest energy select the final state. Specifically, we relax translational symmetry in the mean-field ansatz by promoting the uniform parameter $A$ to a site-dependent $A_i$, initialized as $A_i = \pm 1$ depending on the seeded ground/Rydberg state. We then compare the energies obtained from the VBS and SS initializations with those from a symmetric random start. We choose two initialization patterns: the valence bond solid (VBS) with Rydberg density $n_{Ryd}=1/4$ and the stripe solid (SS) with $n_{Ryd} = 1/3$ [Fig.~\ref{fig:GS}(a)]. The VBS spontaneously breaks the rotation and translation symmetries, doubling the unit vectors in both directions. In contrast, the SS phase breaks translation symmetry along only one axis, concentrating excitations along parallel lines. This structure arises because SS necessarily violates the inter-triangle Rydberg blockade; however, due to the $1/r^6$ interaction, $\raisebox{-0.32\height}{\includegraphics[height=1.3em]{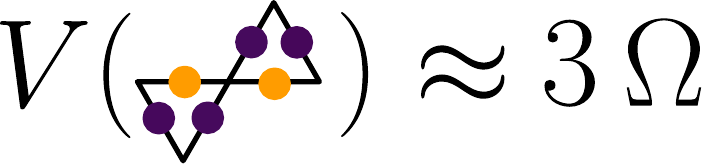}}$ is energetically less costly than $\raisebox{-0.32\height}{\includegraphics[height=1.3em]{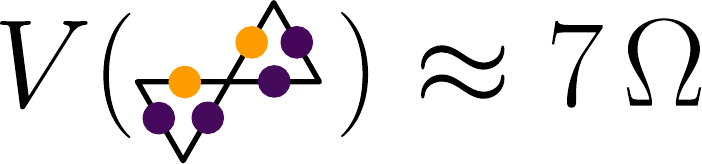}}$, making aligned Rydberg excitations more favorable. In Fig.~\ref{fig:GS}(b), we observe that for sufficiently large $\delta$, the solid-initialized state attains lower energy, indicating a transition to a solid ground state.

    To distinguish between different symmetry-breaking patterns, we define the correlation:
    \begin{equation}\label{eq:order_params}
    \begin{split}
        &\mathcal{M}_\text{VBS} = {N^{-1}_\text{cell}} \sum_{n_1, n_2} (-1)^{n_1+n_2} u_{n_1, n_2} \\
        & \mathcal{M}_\text{SS} = {N^{-1}_\text{cell}} \sum_{n_1, n_2} (-1)^{n_1} u_{n_1, n_2},
    \end{split}
    \end{equation}
    where $n_1$ and $n_2$ label unit cell positions as a multiple of the respective unit vectors $\mathbf{a}_1$ and $\mathbf{a_2}$ with $\mathbf{a}_2$ being the symmetric direction of the SS configuration. $u_{n_1, n_2}$ is a binary variable which depends on the state $s_{n_1, n_2}$ of the up-triangle in the unit cell at $(n_1, n_2)$. Specifically, 
    \begin{equation}
        u_{n_1, n_2} = \begin{cases} s_{n_1, n_2} \circ s_{0, 0} &\text{ for even }n_1\\
        \left( s_{n_1, n_2} \circ s_{1, 0} \right)(s_{1, 0} \circ s_{0,0}) &\text{ for odd }n_1
        \end{cases},
    \end{equation}
    where $s\circ s'=1$ if $s=s'$ and $-1$ otherwise. With this choice of $u_{n_1, n_2}$, the VBS and SS order parameters are maximized for the checkerboard and stripe patterns, respectively. As shown in Fig.~\ref{fig:GS}(c), the order parameters $\mathcal{M}_\text{VBS}$ and $\mathcal{M}_\text{SS}$ become finite in their respective phases and diminish elsewhere, establishing two phase transitions: trivial-VBS at $\delta/\Omega \approx 3.4$ and VBS-SS at $\delta/\Omega\approx 5.0$. In the language of $e$ and $m$ anyons defined as violations of terms in Eqn.~\eqref{eq:stabilizerlimit}, the trivial phase is an $e-$condensate where local Gauss laws are not enforced; while the VBS phase is an $m-$condensate with observed Gauss law but lacking resonance.

    \begin{figure}[t]
		\centering
		\includegraphics[width=0.3\textwidth]{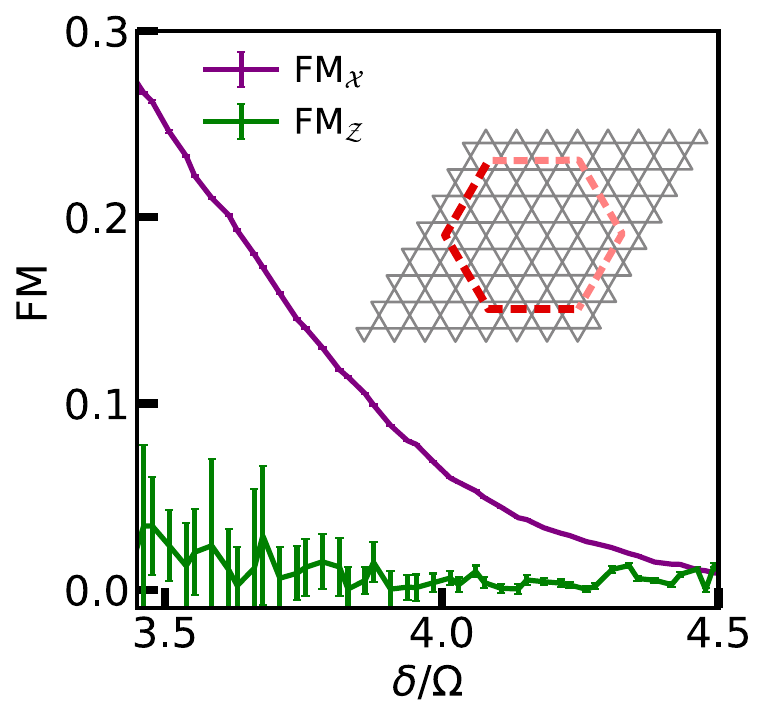}
		\caption{FM order parameter with fixed $\Gamma\approx 0.096~\Omega$. Both the diagonal and off-diagonal FM order parameters approach zero as $\delta$ is swept into the VBS phase. The inset depicts the closed loop, and the open string corresponding to half that loop, used for computing the FM order parameters.}
		\label{fig:FM}
    \end{figure}

    \emph{String order parameter}---In addition to the TEE over a finite subsystem, here we provide another finite-size metric for the topological order, namely the Fredenhagen-Marcu (FM) order parameter \cite{Gregor2011, xu2025critical}. The diagonal FM$_\mathcal{Z}$ characterizes the confinement of $m$ excitations,
    \begin{equation}
        \text{FM}_\mathcal{Z} = \frac{\langle \psi | \Pi_{l\in C'} \mathcal{Z}_l | \psi \rangle}{\sqrt{\langle \psi | \Pi_{l\in C} \mathcal{Z}_l | \psi \rangle}},
    \end{equation}
    where $C$ are closed loops, and $C'$ are open strings corresponding to half the respective closed loops [see inset of Fig.~\ref{fig:FM}]. The off-diagonal FM$_\mathcal{X}$ is defined similarly but with the elementary $\mathcal{X}$ instead of $\mathcal{Z}$, and diagnoses the confinement of $e$ anyons.  The vanishing of both FM order parameters is expected in a quantum spin liquid, and this is indeed observed for the optimal ramp (Fig.~\ref{fig:FM}). Together with the convergence of the TEE to values close to $\gamma=\ln 2$ [Fig.~\ref{fig:drawing_3}(b)], we establish the spin liquid-like signatures near the optimal ramping rate up to the length scale accessible in our numerics. 
    \clearpage
    
	\appendix

    \clearpage
    \clearpage\onecolumngrid
    \section*{Supplementary Material \label{sec:SM}} 
    
    \setcounter{figure}{0}
    \setcounter{table}{0}
    \setcounter{equation}{0}
    \setcounter{section}{0}
    
    \renewcommand{\thefigure}{S\arabic{figure}}
    \renewcommand{\thetable}{S\arabic{table}}
    \renewcommand{\theequation}{S\arabic{equation}}
    \renewcommand{\thesection}{S\arabic{section}}
    
    \makeatletter
    \renewcommand{\theHfigure}{S\arabic{figure}}
    \renewcommand{\theHtable}{S\arabic{table}}
    \renewcommand{\theHequation}{S\arabic{equation}}
    \renewcommand{\theHsection}{S\arabic{section}}
    \makeatother

	\section{Numerical methods}\label{ap:numerical}
    In this section, we detail the numerical methods used in the main text. We first describe the implementation of the approximately gauge-symmetric architecture on the ruby lattice. Unlike the case of the toric code ground states studied in Ref. \cite{kufel2025}, simulating Rydberg-atom dynamics on the ruby lattice is more susceptible to numerical instabilities and error accumulation. To address these challenges, we employ several new techniques, including an auxiliary mean-field ansatz, update-clip regularization, an adaptive sampling transition rule, and ratio estimators for R\'enyi entropy computation.
    
	\subsubsection{Approximately-gauge symmetric neural ansatz}
    We start by discussing the reduction of the effective Hilbert space due to the Rydberg blockade. As mentioned in the main text, for experimental parameters $R_b = 2.4a$, the strong blockade within each triangle ($V_{ij}/\Omega \sim \mathcal{O}(10^2)$) makes configurations with more than one excitation per triangle energetically costly and therefore strongly suppressed. We therefore restrict the local Hilbert space on each triangle to the four states with at most a single excitation (effective spin-$3/2$). We encode these 4 possible states by a binary 2-element column:
	\begin{equation}
		\raisebox{-25pt}{\includegraphics[width=0.4\linewidth]{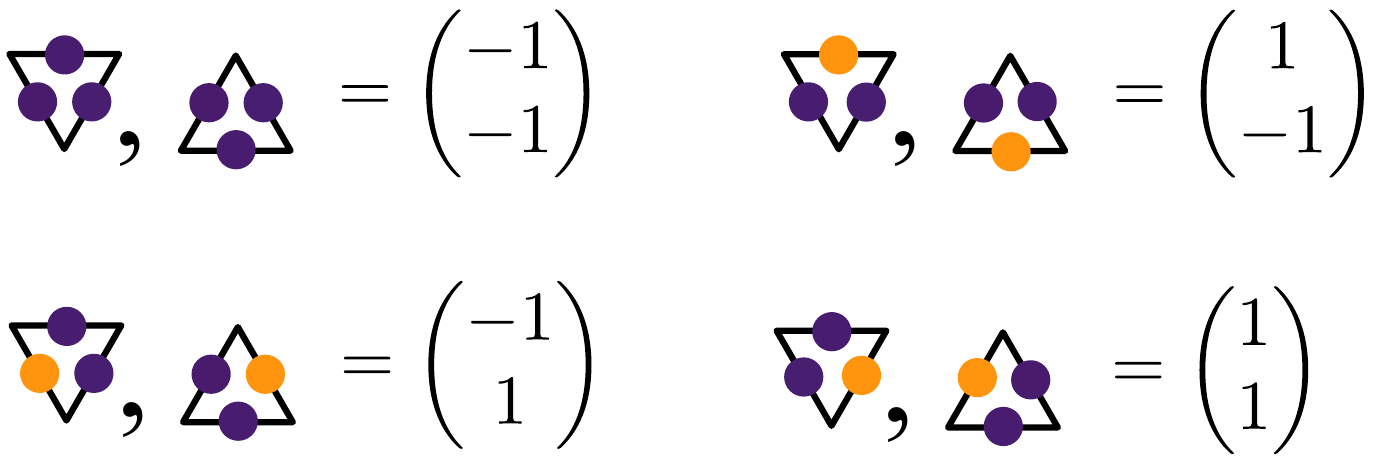}}.
        \label{eq:s1}
	\end{equation}   
    We emphasize that restricting the local Hilbert space does not preclude two excited atoms within the Rydberg blockade radius if they occupy adjacent triangles. We can further define the $\mathbb{Z}_2$ parity charge at each vertex by the following
	\begin{equation}
		\raisebox{-10pt}{\includegraphics[width=0.3\linewidth]{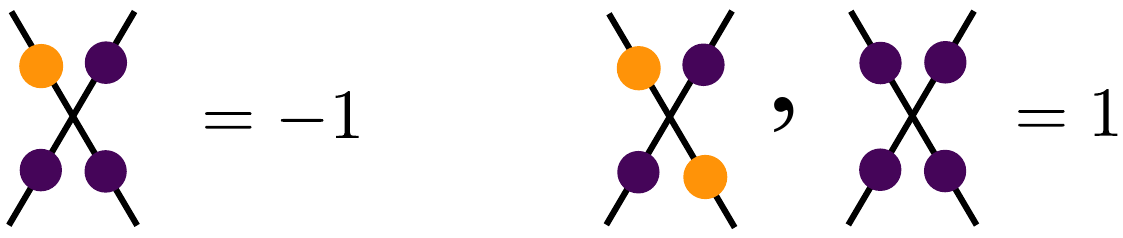}}.
        \label{eq:s2}
	\end{equation}

We represent the logarithm of the variational wavefunction, $\log \psi_\theta(s)$, by complex-valued convolutional neural networks arranged into two parallel branches: an explicitly gauge-invariant branch and a gauge-non-invariant branch. The first branch processes the $\mathbb{Z}_2$ charge configurations given by Eqn.~\ref{eq:s2}, ensuring that all intermediate features and the output are exactly gauge-symmetric. The other branch takes the excitation configuration as input and thus does not enforce any symmetry. This composite construction follows the general philosophy of approximately symmetric neural quantum states, in which one combines an exactly symmetric component with a more flexible component rather than enforcing the symmetry at every stage of the network.

To exploit lattice translation symmetry, we first group the microscopic degrees of freedom within each unit cell into a single \emph{super-site}. After this reshaping step, the lattice is treated as an $L_1 \times L_2$ array of $N_{\mathrm{hex}} = L_1 L_2$ super-sites, one per physical hexagon. For the non-invariant branch, the local input feature vector at each super-site contains the four microscopic variables (two from the encoding Eqn.~\ref{eq:s1} and another factor two from the up and down-triangle sublattices) associated with that unit cell, so that the input tensor has shape
\begin{equation}
\boldsymbol{u}^{(0),\mathrm{non\text{-}inv}} \in \{1, -1\}^{L_1 \times L_2 \times 4}.
\end{equation}
That is, the input is a rank-3 tensor with two spatial indices labeling the unit cell and one channel index labeling the four internal degrees of freedom.

For the invariant branch, the microscopic variables are first mapped to a gauge-invariant representation using the local gauge-invariant product nonlinearity defined in Eq.~\ref{eq:s2}. Concretely, for each unit cell, we evaluate the three vertex-centered parities associated with the kagome vertices belonging to that hexagon. This produces an input tensor
\begin{equation}
\boldsymbol{u}^{(0),\mathrm{inv}} \in \{1, -1\}^{L_1 \times L_2 \times 3}.
\end{equation}
Thus, the invariant branch does not receive the microscopic variables directly, but only composite variables that already lie in the gauge-invariant sector.

Each branch is processed by a stack of convolutional layers. If $\boldsymbol{u}^{(n-1)} \in \mathbb{C}^{L_1 \times L_2 \times d^{(n-1)}}$ denotes the feature tensor entering layer $n$, the output is
\begin{equation}
\boldsymbol{u}^{(n)} =
\sigma_n \left(
\boldsymbol{W}^{(n-1)} * \boldsymbol{u}^{(n-1)} + \boldsymbol{b}^{(n-1)}
\right),
\end{equation}
where $d^{(n-1)}$ is the number of channels in layer $n-1$, $\sigma_n$ is the activation function used in layer $n$, and the discrete convolution is defined by
\begin{equation}
\left[\boldsymbol{W}^{(n-1)} * \boldsymbol{u}^{(n-1)}\right]_{\boldsymbol{a},i}
=
\sum_{j=1}^{d^{(n-1)}} \sum_{\boldsymbol{r} \in \mathcal{K}}
\boldsymbol{W}^{(n-1)}_{\boldsymbol{r},ij}
\boldsymbol{u}^{(n-1)}_{\boldsymbol{a}-\boldsymbol{r},j}.
\label{eq:convolve_appsym}
\end{equation}
Here $\boldsymbol{a} = a_1 \boldsymbol{a}_1 + a_2 \boldsymbol{a}_2$ labels the unit cell, $i$ is the output-channel index, and $\mathcal{K}$ is the set of lattice displacements contained in the convolution kernel. In implementation, the displacement is limited to a rectangle of size $(R_1, R_2)$ with periodic padding. The trainable parameters are the complex kernel weights $\boldsymbol{W}^{(n-1)}$ and biases $\boldsymbol{b}^{(n-1)}$.

In our implementation, both branches contain three hidden convolutional layers with channel dimensions
\begin{equation}
d^{(1)} = d^{(2)} = d^{(3)} = 8
\end{equation}
The kernel linear size is chosen as $l \sim L/N_h$, where $L$ is the linear system size and $N_h = 3$ is the number of hidden layers. Operationally, we use square kernels of size $l \times l$ with odd $l$, centered on the origin, so that
\begin{equation}
\mathcal{K} =
\left\{
-\frac{l-1}{2}, \ldots, \frac{l-1}{2}
\right\}^2.
\end{equation}
With this choice, after $N_h$ layers information can propagate across the full system.

After the last hidden layer, the two branches are recombined additively at each spatial position,
\begin{equation}
\boldsymbol{v}
=
\boldsymbol{W}^{(N_h-1),\mathrm{inv}} * \boldsymbol{u}^{(N_h-1),\mathrm{inv}}
+
\boldsymbol{W}^{(N_h-1),\mathrm{non\text{-}inv}} * \boldsymbol{u}^{(N_h-1),\mathrm{non\text{-}inv}}
+
\boldsymbol{b}^{(N_h-1)}.
\end{equation}
We then apply a final nonlinearity and average over all spatial positions and output channels,
\begin{equation}
\theta^{\mathrm{NN}}
=
\mathrm{AvgPool}\!\left[
\sigma_{\mathrm{out}}(\boldsymbol{v})
\right].
\end{equation}
The average pooling is taken uniformly over all $L_1 \times L_2$ unit cells and over the final channel indices, yielding a single complex scalar.

Because the variational wavefunction is complex-valued, we choose activation functions that are holomorphic polynomials rather than non-analytic functions. For the first two hidden layers we use
\begin{equation}
\sigma_1(x) = \sigma_2(x) = \frac{x}{2} + \frac{x^2}{4},
\end{equation}
while for the final layer before pooling we use
\begin{equation}
\sigma_{\mathrm{out}}(x) = \frac{x}{2} + \frac{x^2}{4} - \frac{x^4}{48}.
\end{equation}
These are low-order Taylor approximations to the swish activation function, $\sigma(x)=x/(1+e^{-x})$. We use polynomial approximations rather than the exact swish function in order to preserve holomorphicity for complex inputs.

To include other lattice symmetries such as the $C_6$ rotation and the $\mathcal{M}_{x, y}$ mirror reflections, we symmetrize the output by
	\begin{equation}
		\tilde\theta^{NN}(s) = \frac{1}{|G|}\sum_{g\in G} \theta^{NN}(g s).	
	\end{equation} 
We note that due to the way we encode the physical lattice, a physical transformation not only permutes elements of the inputs but also changes their values. For example, a reflection changes the position of a triangle and additionally exchanges the internal states $\raisebox{-0.32\height}{\includegraphics[height=1.3em]{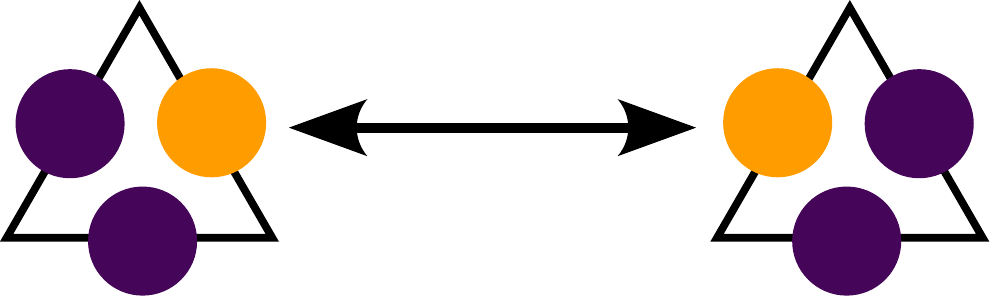}}$.

    Next, to enhance stability during time evolution for dynamics or optimization for ground-state search, we insert a mean-field ansatz beside the symmetrized neural networks [see End Matter of the main text]
    \begin{equation}
        \log \psi(s) = \theta^{MF}(s) + \tilde{\theta}^{NN}(s).
    \end{equation}
    We note that the mean-field ansatz used here benefits both the ground-state optimization and dynamics simulations performed in the main text, especially at early times. In the variational Monte Carlo calculation, $ \log \psi(s)$ defines the logarithm of the wavefunction amplitude for the sampled configuration. Finally, we further test convergence on large systems by confirming that the dynamics of observables are unchanged across several choices of hyperparameters.
    
    \subsubsection{Time-evolution and TDVP}

    \begin{figure}
		\centering
		\includegraphics[width=0.4\linewidth]{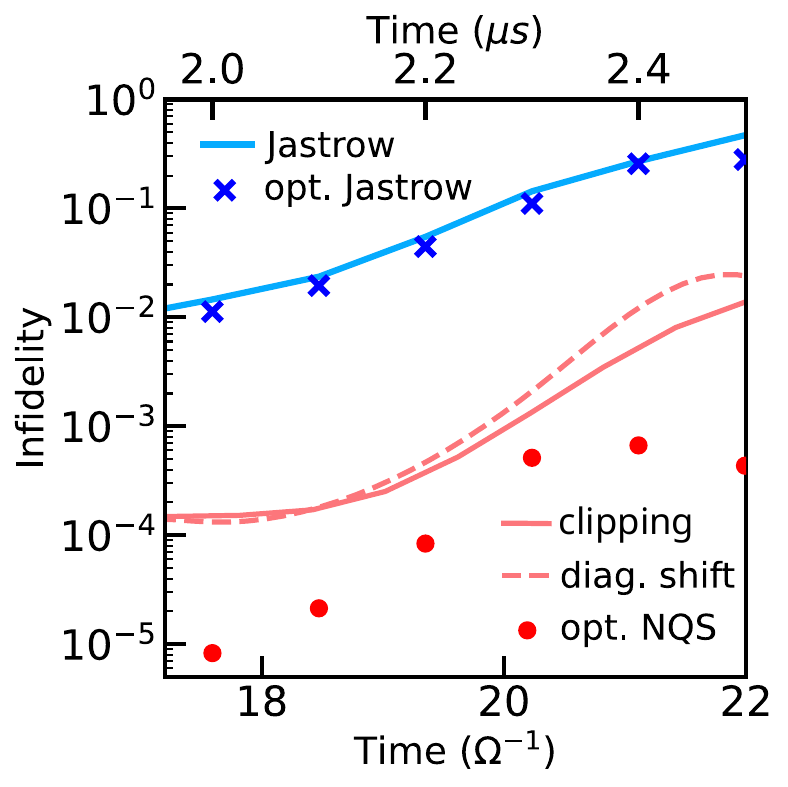}
        \caption{Infidelity of simulated dynamics compared with exact diagonalization for an $N=24$ system under the ramping protocol~\cite{semeghini2021probing}. Lines show TDVP results; points indicate infidelity from direct optimization with the exact ED wavefunction. Blue lines and crosses: decorated Jastrow ansatz (data from \cite{mauron2025predicting}). Dashed red line: TDVP with approximately symmetric NQS and diagonal-shift regularization ($\epsilon=10^{-6}$). Solid red line and dots: TDVP with approximately symmetric NQS using update-clip regularization. \label{fig:sup_infidelity}}
	\end{figure}
    
    The NQS at the beginning of the ramp is obtained by optimizing the energy $\bra{\psi}H(t=0)\ket{\psi}$. From the initial state, we evolve the neural-network state in time using the time-dependent variational principle (TDVP), which governs the evolution of the network parameters $\boldsymbol{\alpha}(t)$. TDVP can be understood as finding a parameter update $\left(\delta \boldsymbol{\alpha}\right)^*$ such that
	\begin{equation}\label{eq:Fubini}
		\left(\delta \boldsymbol{\alpha}\right)^*=\textrm{argmin}_{\delta \boldsymbol{\alpha}} \mathcal{D}(\ket{\psi_{\boldsymbol{\alpha} + \delta \boldsymbol{\alpha}}}, e^{-iH\tau} \ket{\psi_{\boldsymbol{\alpha}}} ), 
	\end{equation}
	where one chooses the quantum state infidelity as a measure of distance, 
	\begin{equation}
		\mathcal{D}(|\phi\rangle,|\psi\rangle) = 1 - \sqrt{\frac{\langle\phi|\psi\rangle\langle\psi|\phi\rangle}{\langle\phi|\phi\rangle\,\langle\psi|\psi\rangle}}. 
	\end{equation}
    One way of arriving at a solution to this equation is by direct minimization of this distance function, which can be accomplished, e.g., by (natural) gradient descent \cite{Sinibaldi2023ptVMC, gravina2024ptVMC}. 
    
    Alternatively, when $\boldsymbol{\delta \alpha} = \boldsymbol{\dot  \alpha} \tau$ and $\tau\ll1$ we can expand Eq.~\eqref{eq:Fubini} up to the second order in $\mathcal{O}(\tau^2)$. The quadratic function can be minimized analytically and gives a closed-form solution for the parameter update equation \cite{carleo2017solving,sorella1998green}
	\begin{equation}\label{eq:tdvp}
		\sum_{k}\boldsymbol{S}_{jk}\dot{\alpha}_k = -i\boldsymbol{F}_j
	\end{equation} 
	where the quantum geometric tensor $\boldsymbol{S}$ and the force $\boldsymbol{F}$ are given by
	\begin{equation}\label{eq:SandF}
		\begin{split}
			&\boldsymbol{S}_{jk} = \frac{\braket{\partial_{\boldsymbol{\alpha_j}}\psi}{\partial_{\boldsymbol{\alpha_k}}\psi}}{\braket{\psi}{\psi}} - \frac{\braket{\partial_{\boldsymbol{\alpha_j}}\psi}{\psi}}{\braket{\psi}{\psi}} \frac{\braket{\psi}{\partial_{\boldsymbol{\alpha_k}}\psi}}{\braket{\psi}{\psi}}, \\
			&\boldsymbol{F}_j =  \frac{\braket{\partial_{\boldsymbol{\alpha_j}}\psi}{H\psi}}{\braket{\psi}{\psi}} - \frac{\braket{\partial_{\boldsymbol{\alpha_j}}\psi}{\psi}}{\braket{\psi}{\psi}} \frac{\braket{\psi}{H\psi}}{\braket{\psi}{\psi}}.
		\end{split}
	\end{equation}
    The system of equations~\eqref{eq:tdvp} is the core of the TDVP method, which reduces quantum time evolution to iteratively solving a set of coupled ODEs for $\dot{\boldsymbol{\alpha}}$. The resulting one-step update is integrated using the second-order Heun method. We choose the time step $dt=10^{-2}~\Omega^{-1}$. We note that by replacing $-iF_j\to-F_j$, Eqn.~\ref{eq:tdvp} describes imaginary-time evolution and can be used for ground-state optimization.

    \subsubsection{TDVP regularization}

    Naively solving the TDVP equation is challenging because $S$ is typically ill-conditioned, making its (pseudo-) inversion numerically unstable. A common remedy is to apply Tikhonov regularization \cite{tikhonov1977solutions}, which adds a small constant to the diagonal elements, $S_{ii} \to S_{ii} + \epsilon$. This shifts all eigenvalues, $\lambda_i \to \lambda_i + \epsilon$, predominantly affecting the smallest ones and effectively capping the inverse eigenvalues at $1/\epsilon$. However, choosing a large $\epsilon$ can distort the trajectory of $\boldsymbol{\alpha}(t)$, deviating it from the true real-time evolution. 
    
    A related approach is to apply an eigenvalue cutoff for the $S$-matrix. It can be understood as follows. Diagonalizing
    \begin{equation}
    \boldsymbol{S} = \boldsymbol{U} \boldsymbol{\Lambda} \boldsymbol{U^{\dagger}},
    \end{equation}
	    where $U$ is unitary and $\lambda$ is a diagonal matrix of singular values, Eq.~\eqref{eq:tdvp} can be written in the eigenbasis as
	\begin{equation}
		\boldsymbol{\tilde{\dot \alpha}} = \boldsymbol{\Lambda}^{-1} \boldsymbol{\tilde{F}}.
	\end{equation}
    with $\boldsymbol{\tilde{\dot \alpha}} = \boldsymbol{U^{\dagger}} \boldsymbol{\dot{\alpha}}$ and  $\tilde{\boldsymbol{F}} = -i \boldsymbol{U}^{\dagger} \boldsymbol{F}$. Within this regularization method, eigenvalues smaller than a chosen threshold, $\lambda < \lambda_{\text{threshold}}$, are discarded.

    In this work, we use a different approach: instead of regularizing the singular values $\lambda_k$ of the matrix $S$, we focus on $\tilde{\dot \alpha}_k = \tilde{F}_k/\lambda_k$. Specifically, $\tilde{\dot \alpha}_k$ is clipped at $\mu_k$ by   
\begin{equation}
\tilde{\dot{\alpha}}_k \;\to\;
\begin{cases}
\tilde{\dot{\alpha}}_k, & |\tilde{\dot{\alpha}}_k| \le \mu_k, \\[6pt]
\mu_k\,\dfrac{\tilde{\dot{\alpha}}_k}{|\tilde{\dot{\alpha}}_k|}, & |\tilde{\dot{\alpha}}_k| > \mu_k~.
\end{cases}
\end{equation}
	where we define
	\begin{equation}
		\mu_k = \begin{cases}
			+\infty & \text{if $\lambda_k /\max_{k}(\lambda) > 10^{-5}$ }\\
			0.5 & \text{if $10^{-5} \ge \lambda_k /\max_{k}(\lambda) > 10^{-8}$} \\
			0.01 & \text{otherwise}
		\end{cases}.
        \label{eq:regularization}
	\end{equation}
    
    Compared to diagonal-shift regularization, which constrains the \emph{overall} magnitude of the update vector $\sum_j |\tilde{\dot{\alpha}}_j|^2$, our approach imposes \emph{individual bounds} on each coordinate. This per-coordinate clipping directly suppresses excessively large components $\tilde{\dot{\alpha}}_k$ that typically arise from small singular values $\lambda_k$ in the quantum geometric tensor. By selectively damping these ill-conditioned directions, our method allocates the numerical error budget more efficiently. From the perspective of eigenvalue regularization, it can be viewed as a softer variant that lowers the noise floor by removing the most destabilizing components.

    To evaluate the effectiveness of our regularization strategy, we compare the per-coordinate constraint with the conventional diagonal shift (global constraint). As shown in Fig.~\ref{fig:sup_infidelity}, our approach reduces the final-state error by roughly a factor of two, and this improvement remains robust to small variations in the regularization parameters in Eq.~\eqref{eq:regularization}. For highly expressive NQS architectures, directly optimizing the overlap with the exact wavefunction achieves significantly higher accuracy than TDVP-based evolution, underscoring the importance of maintaining numerical stability during time evolution (and consequently of the regularization method). In contrast, for the more restricted decorated Jastrow ansatz \cite{mauron2025predicting}, the TDVP results already reach the accuracy limit set by direct optimization, rendering further fine-tuning unnecessary.

    \subsubsection{Sampling and Markov Chain Monte Carlo statistics}

	\begin{figure}
		\centering
		\includegraphics[width=0.7\linewidth]{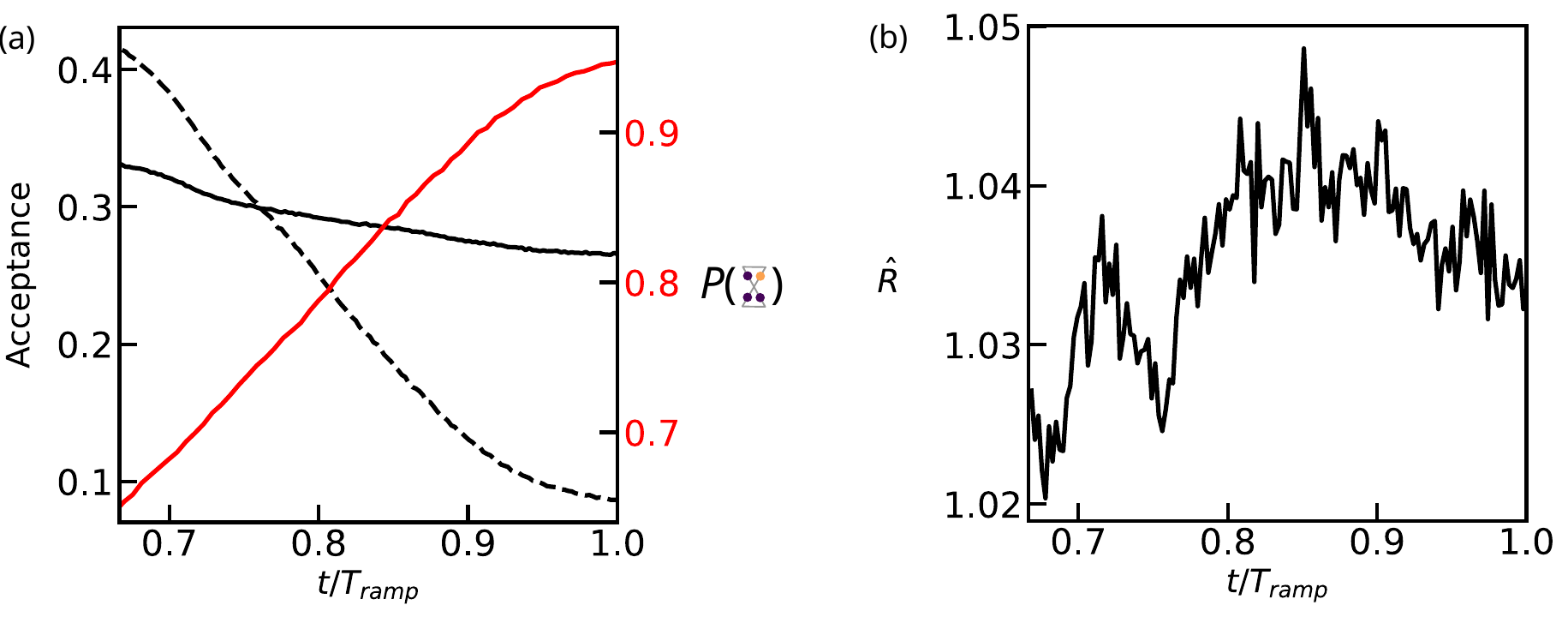}
			\caption{(a) Markov-chain acceptance rate with plaquette-flip probability $p=0.1$ (dashed) and $p=0.4$ (solid). The red line shows the projection of the state onto the Gauss-law-satisfying manifold. Single-spin flips yield lower acceptance when the state converges to a small sector within the Hilbert space. (b) Split-Rhat diagnostic $\hat{R}$ from Hamiltonian evaluation during the ramp. Good simulations have $\hat{R}$ values less than $1.1$ \cite{vehtari2021rank}.}\label{fig:diagnosis}
	\end{figure}

    We now describe the calculation of observables within the NQS ansatz, evaluated using Markov chain Monte Carlo sampling. Multiple Markov chains of bit-string configurations, $s_0 \rightarrow s_1 \rightarrow \dots$, are generated via the Metropolis–Hastings algorithm with a chosen update rule. The standard NQS update involves single-spin flips; however, at late ramp times, when Gauss laws emerge, this approach leads to low acceptance rates and poor ergodicity (see Fig.~\ref{fig:diagnosis}). To improve sampling, we adopt a hybrid update scheme combining single-spin and plaquette flips. The plaquette flips, analogous to the action of $\mathcal{X}_6$, preserve the Gauss laws. During each Markov step, a plaquette flip is chosen with probability $p$ and a single-spin flip with probability $1-p$. Convergence across chains is monitored using the split-$\hat{R}$ diagnostic, with $\hat{R}<1.1$ indicating good convergence. Setting $p=0.4$ near the end of the ramp yields an acceptance rate above 20\% while maintaining $\hat{R}<1.05$.
    	
	\subsubsection{R\'enyi$-2$ entropy computation}
	\begin{figure}
		\centering
		\includegraphics[width=0.6\linewidth]{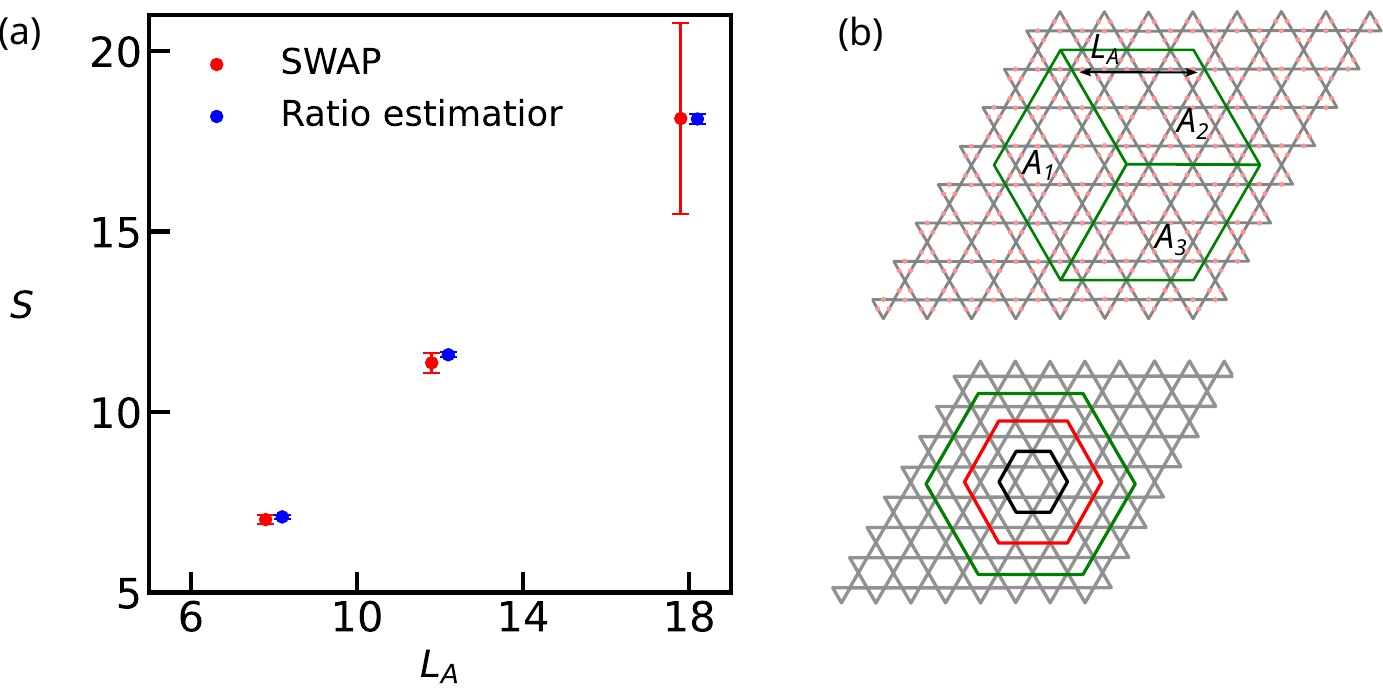}
	        \caption{(a) R\'enyi entropy with respect to the subsystem boundary lengths evaluated by the naive implementation of the SWAP operator (red) and the sequential ratio estimators (blue) at the end of the ramp with $\Gamma = 0.096~\Omega$. The ratio estimators use 48 intermediate points between $t=0.8~T_{ramp}$ and $t=T_{ramp}$, with the number of samples $N_s=294912$ in both cases. (b) Top: Division of a subsystem (the largest hexagon) into three adjacent partitions for the computation of TEE according to Eq.~\eqref{eq:Kitaev_Preskill}. The depicted subsystem has linear size $L_A = 6$. Bottom: Subsystem sizes to study the scaling of TEE. The colors correspond to Fig.~3 of the main text with $L_A = 2, 4, 6$ from the smallest to largest.}
		\label{fig:partition}
	\end{figure}

	    Next, we discuss how to evaluate the entanglement entropy within variational Monte Carlo. The entanglement entropy is non-linear in the density matrix, but can be accessed by measuring an expectation value on two copies
	\begin{equation}\label{eq:entropy}
		S_A = -\ln \text{Tr}\left(\rho_A^2\right) = -\ln \text{Tr}\left[(\rho \otimes \rho) \text{SWAP}_A \right],
	\end{equation} 
	where $\rho_A= \text{Tr}_{\bar{A}}\rho$ is the reduced density matrix obtained by tracing out the complementary partition $\bar{A}$, $\rho \otimes \rho$ is built from stacking two copies of the state $\rho$ and $\text{SWAP}_A$ is the operator swapping all sites within subsystem $A$ between the two copies. Eq.~\eqref{eq:entropy} provides a route to evaluate $S$ by sampling over the doubled Hilbert space. A basis element $x$ of the doubled Hilbert space refers to the concatenation $(\sigma, \eta)$ of two basis vectors in the original Hilbert space. For conciseness, we denote the integrand of the SWAP action as
    \begin{equation}
        F(x; A, \psi) = \psi(\eta_{A}, \eta_{\bar{A}}) \psi(\sigma_{A}, \sigma_{\bar{A}}) \psi^*(\eta_{A}, \sigma_{\bar{A}}) \psi^*(\sigma_{A}, \eta_{\bar{A}}).
    \end{equation}
    Analogously, $F(x; \emptyset, \psi) = |\psi(\sigma)|^2 |\psi(\eta)|^2$ and 
    \begin{equation}
        S_A = -\log \left[ \frac{\sum_x F(x; A,\psi)}{ \sum_x F(x; \emptyset,\psi)} \right] = -\log \left[ \mathbb{E}_{x\sim \mathbb{P}(x)} \frac{F(x; A, \psi)}{F(x; \emptyset, \psi)} \right].
    \end{equation}
    where $\mathbb{P}(x) = F(x; \emptyset,\psi) / \sum_{x'} F(x'; \emptyset,\psi) $ is the distribution in the double Hilbert space. Since we use Markov chain Monte Carlo sampling, we draw samples from the unnormalized $F(x; \emptyset,\psi)$.
    
    In general, the entanglement entropy scales with either the boundary or the volume of the subsystem, making $\langle \text{SWAP} \rangle$ exponentially small in system size and thus often requiring an impractically large number of samples. To overcome the sampling issue, Ref.~\cite{hastings2010measuring} proposed the ``ratio estimator'' route by computing the entropy increment along a path. Specifically, 
    \begin{equation}
        S_{A_N} = (S_{A_N} - S_{A_{N-1}}) + (S_{A_{N-1}} - S_{A_{N-2}}) + \dots+ (S_{A_1} - S_{A_0}) + S_{A_0} =  S_{A_0}  + \sum_n \Delta S_n 
    \end{equation}
    where $\{A_0, A_1, \dots, A_N \}$ is a sequence of larger partitions from the smallest $A_0$ to the target partition $A_N$. Since entropy scales with the subsystem size, one can, in principle, choose a sequence such that $\Delta S_n
    \approx$~const. Then, by evaluating the individual $\Delta S_n$ and summing them up, one can shift the exponential cost $e^{S}$ to a linear cost, $(S / \Delta S) e^{\Delta S}$. The key advantage is that one can evaluate $\Delta S$ directly by reweighting sampling. The entropy difference, e.g. $\Delta S = S_{A_1} - S_{A_0}$ is estimated by
    \begin{equation}\label{eq:delta_S}
        \Delta S = -\log \left[ \frac{\sum_x F(x; A_1,\psi)}{ \sum_x F(x; A_0,\psi)} \right] = -\log \left[ \mathbb{E}_{x\sim F(x; A_0, \psi)} \frac{F(x; A_1, \psi)}{F(x; A_0, \psi)} \right].
    \end{equation}
    We note that $F(x; A_0,\psi)$ has to be non-negative to become a valid distribution. This can be satisfied automatically if $\psi$ is real and non-negative.

    To adapt this idea to our case, we note that (i) the wavefunction is complex and (ii) the path we use to compute entropy increments is controlled by the evolution time instead of scaling the partition size. To tackle complex wavefunctions, we first distinguish the amplitude and phase, $\psi(\sigma) = \Psi(\sigma)e^{i\theta(\sigma)}$. Then the total entropy $S = S^\text{Am} + S^\text{Ph}$, where
    \begin{equation}
        S^\text{Am} = -\log \left[ \frac{\sum_x F(x; A,\Psi)}{\sum_x F(x; \emptyset,\Psi)} \right],\quad S^\text{Ph} = -\log \left[ \frac{\sum_x F(x; A,\psi)}{\sum_x F(x; A,\Psi)} \right].
    \end{equation}
    By construction $F(x; A, \Psi)$ is non-negative as $\Psi$ is non-negative, so $S^\text{Ph}$ can be evaluated using the same framework as Eq.~\eqref{eq:delta_S}. Under a slow ramp, the state stays close to a ground state that is nonnegative, so $S^\text{Ph}$ is usually small, and does not suffer from the exponential complexity. For the TEE shown in the main text, we include both $S^\text{Am}$ and $S^\text{Ph}$.
    
    The extensiveness of entropy is mostly contained in $S^\text{Am}$, which can be handled by sufficient ratio estimators at intermediate steps. The dynamics of interest starts with a product state with small entanglement entropy that can be efficiently evaluated by the usual naive method. Additionally, the entropy increment from one instant to another can be made arbitrarily small by tuning the time step, allowing us to evaluate the entanglement entropy and topological entanglement entropy to high resolution. Specifically, $\Delta S_t = S^\text{Am}_t - S^\text{Am}_{t-\tau}$ can be evaluated via
    \begin{equation}\label{eq:delta_St}
        \Delta S_t =  -\log \left[ \frac{\sum_x F(x; A,\Psi_t)}{\sum_x F(x; A,\Psi_{t-\tau})} \right] + \log \left[ \frac{\sum_x F(x; \emptyset,\Psi_t)}{\sum_x F(x; \emptyset, \Psi_{t-\tau})} \right].
    \end{equation}
    Compared to Eq.~\eqref{eq:delta_S}, Eq.~\eqref{eq:delta_St} contains two terms, but the extra term does not depend on the subsystem and can be reused for other subsystems.
    
	To obtain the topological entanglement entropy over a subsystem $A = A_1 \cup A_2 \cup A_3$ [see Fig.~\ref{fig:partition}(b)], we use the Kitaev-Preskill formula \cite{kitaev2006tee}
	\begin{equation}\label{eq:Kitaev_Preskill}
		-\gamma = S_{A_1} + S_{A_2} + S_{A_3} - S_{A_1A_2} - S_{A_2A_3} - S_{A_1A_3} + S_{A_1A_2A_3}.
	\end{equation}
	By scaling the size of $A$, we reveal the finite-size characteristics of the dynamically prepared QSL.

    \section{Additional numerical data} \label{ap:more_data}

	\begin{figure*}
		\centering
		\includegraphics[width=\linewidth]{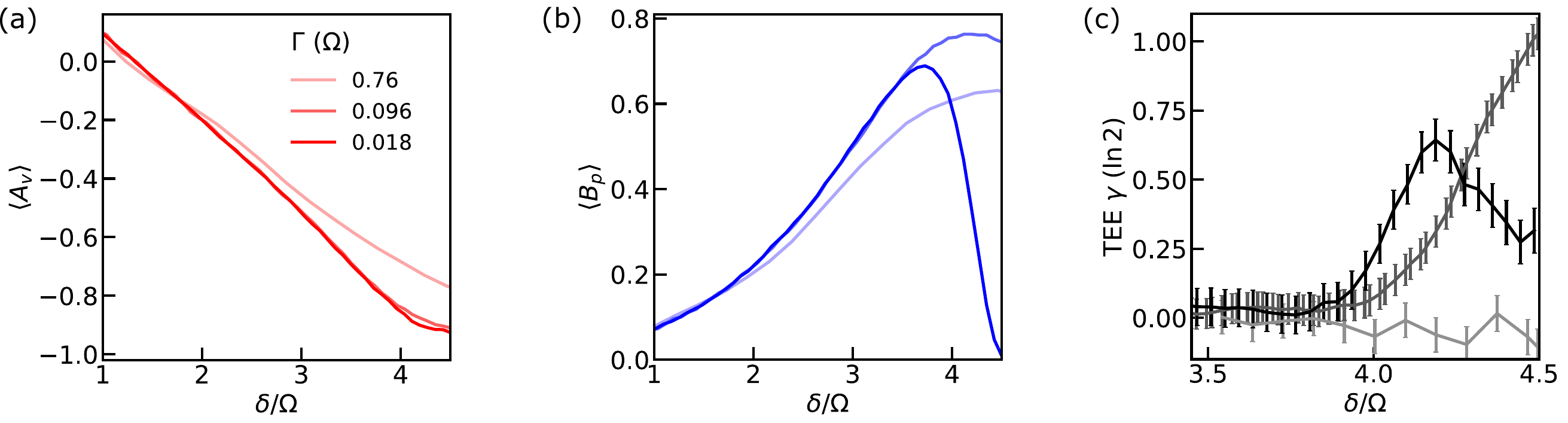}
		\caption{Dynamic preparation under the diabatic, hemi-adiabatic, and adiabatic ramps (from the lightest to boldest colors), tracking various observables: (a) Gauss law, (b) resonance, (c) topological entanglement entropy for $L_A=6$.}
		\label{fig:extra}
	\end{figure*}
    
	    In Fig.~\ref{fig:extra}, we complement Fig.~3 in the main text by comparing the behavior at different ramp rates. In the VBS phase $\delta/\Omega=4.5$, the Gauss' law $\langle A_v\rangle$ moves monotonically closer to its spin-liquid/stabilizer value as the ramp is slowed [Fig.~\ref{fig:extra}(a)]. In contrast, the resonance $\langle B_p \rangle$ exhibits a peak at an intermediate rate, but vanishes when the ramp is closer to the adiabatic limit [Fig.~\ref{fig:extra}(b)]. The difference among diabatic, hemi-adiabatic, and adiabatic ramps is more pronounced in more global properties such as the topological entanglement entropy, as shown in Fig.~\ref{fig:extra}(c). For a fast ramp, the TEE (for $L_A=6$) stays around zero regardless of the detuning, showing that the states at any point remain trivial. In the slow-ramp limit, the state at large $\delta$ also shows vanishing TEE. However, unlike the fast-ramp limit where the zero TEE is the remnant of the initial trivial state, the adiabatically prepared state is close to the VBS ground state of the instantaneous Hamiltonian. This is apparent from a peak in TEE during the sweep, indicating a phase crossover. Therefore, only for intermediate values of the ramping rate does the TEE approach $\ln 2$ as the detuning is swept into the VBS phase.

	In Fig.~\ref{fig:extra_scaling}(a-b), we provide additional data used to extract the two length scales $\lambda_e$ and $\xi_m$ in the main text. The raw data for the Wilson $\mathcal{Z}$ loop are shown in Fig.~\ref{fig:extra_scaling}(a) with respect to the loop perimeter. The data show a deviation from a linear relation at large loops. Indeed, the normalized Wilson-loop measure $W_A^{\mathcal{Z}}/W_A^{\mathcal{Z, gs}}$ exhibits a pure area-law decay, from which we extract the length scale $\lambda$. 

    The length scale $\xi$ used in the main text is extracted from scaling of the open $\mathcal{Z}$ strings with the fitting ansatz
	\begin{equation} \label{eq:xi}
		\langle \mathcal{S}_L^{\mathcal{Z}} \rangle \sim e^{-L/\xi} \times \begin{cases}
			A_1 \text{ if } L - 1 \mod 4 \equiv 0 \\
			A_2 \text{ otherwise } 
		\end{cases}.
	\end{equation}
    The two coefficients $A_1$ and $A_2$ originate from the symmetry breaking pattern in the VBS phase where the ground state unit cell is doubled compared to the ruby lattice unit cell. The data is shown in Fig.~\ref{fig:extra_scaling}(b). This length scale can be related to the perimeter-law decay of Wilson $\mathcal{X}$ loops. $W_A^{\mathcal{X}}$ measures the parity of $\mathcal{Z}$-strings crossing the boundary $\partial A$, thus $ W_A^{\mathcal{X}} \propto e^{-n_m \xi |\partial A|}$. Figure~\ref{fig:extra_scaling}(c) supports our dilute free $e$ gas argument described in the main text by showing that $1/\lambda_e^2 \approx 2(n_e - n_e^{gs})$ at different ramp rates.

    \section{Crossover to perimeter-law scaling}

	\begin{figure*}
		\centering
		\includegraphics[width=\linewidth]{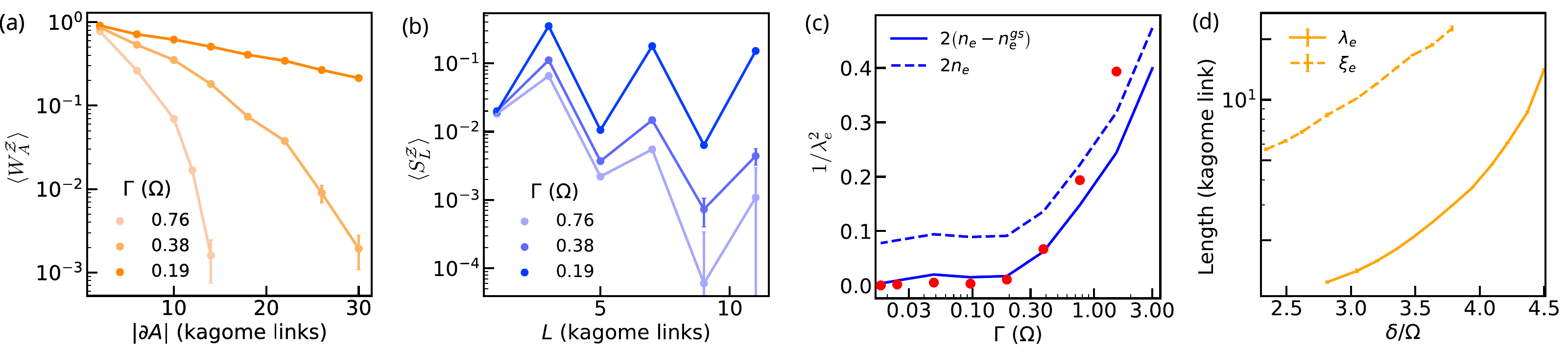}
		\caption{(a) Scaling of the closed $\mathcal{Z}$ loop with respect to the loop perimeter. (b) Amplitude of open $\mathcal{Z}$ strings with respect to the string length. With sufficiently slow ramps, the open strings show prominent solid order. (c) The decay coefficient of the closed $\mathcal{Z}$ loop $1/\lambda_e^2$ with respect to the area (dots) superimposed with the offset (solid line) and unoffset $e$-anyon density (dashed line). The offset is $n_e^{gs}$, the anyon density of the ground state of the final Hamiltonian. (d) The correlation length $\xi_e$ (dashed line) and average distance $\lambda_e$ (solid line) between $e$-excitations for the ramp with $\Gamma = 0.096~\Omega$.}
		\label{fig:extra_scaling}
	\end{figure*}
    
    Even though the area-law scaling of the Wilson $\mathcal{Z}$ loops is well justified within our numerics, it is unstable from the perspective of general gauge theory, i.e., area-law scaling exists only in fine-tuned models with no dynamical matter \cite{FradkinShenker1979}. We now show that by adding correlations to the free $e$-gas picture in the main text, the scaling law crosses over from an area law to a perimeter law when the loop size $R \sim \xi_e$, where $\xi_e$ is the correlation length (to be defined later). We also provide numerical evidence that: (i) the correlation length is much longer than the average $e-e$ distance, justifying the free-gas picture, and (ii) the correlation length exceeds the lattice size, making the ultimate perimeter law inaccessible.
    
    For simplicity, we illustrate the argument using perturbations by $\mathcal{X}$-strings on a clean toric-code wavefunction, which demonstrate the same effect.
    \begin{equation}
    \ket{\psi}
    \propto
    \prod_i
    \left(
        1+\sum_j \alpha_{S_{i:j}} \mathcal{X}_{S_{i:j}}
    \right)
    \ket{\mathrm{TC}} ,
\end{equation}
where $S_{i:j}$ denotes a continuous string beginning at site $i$ and ending at site $j$.
Since the string is physically non-directional, we impose $\alpha_{S_{i:j}} = \alpha_{S_{j:i}}$.
The $e$ excitation density is defined by
\begin{equation}
    2n_e
    =
    1-
    \frac{\braket{\psi|A_i|\psi}}{\braket{\psi|\psi}} .
\end{equation}
The deviation from unity counts the weighted number of $\mathcal{X}$-strings that either start or end at site $i$:
\begin{align}
    2n_e
    =
    2\sum_j |\alpha_{S_{i:j}}|^2
    +
    2\sum_j |\alpha_{S_{j:i}}|^2 =
    4\sum_j |\alpha_{S_{i:j}}|^2 .
\end{align}
Assume a correlation length $\xi_e$ such that $    |\alpha_{S_{i:j}}|^2
    =
    \alpha^2 \exp(-r_{ij}/\xi_e)$.
For simplicity, we approximate the string length by the Euclidean distance $r_{ij}$.
Promoting the discrete sum to a continuum integral gives
\begin{align}
    2n_e
    =
    4\alpha^2
    \int_{\mathbb{R}^2}
    e^{-r/\xi_e}
    \, d^2\mathbf r  
    =
    4\alpha^2
    \int_0^{2\pi} d\theta
    \int_0^\infty r\,dr\, e^{-r/\xi_e}  
    =
    4\alpha^2 \left(2\pi \xi_e^2\right).
\end{align}
Therefore, $\alpha^2
    =
    n_e / (4\pi \xi_e^2).$

Now consider a closed Wilson loop $W_A^\mathcal{Z}$, where $A$ is taken to be a disk of radius $R$.
The loop anticommutes with any string $S_{i:j}$ for which one endpoint lies inside $A$
and the other lies outside $A$. Thus, to leading order in the string density,
\begin{equation}
    \braket{W_A^\mathcal{Z}}
    \approx
    1
    -
    4\alpha^2 \, \mathcal I(R),
\end{equation}
where
\begin{equation}
    \mathcal I(R)
    =
    \int_{|\mathbf r_i|<R} d^2\mathbf r_i
    \int_{|\mathbf r_j|>R} d^2\mathbf r_j\,
    e^{-|\mathbf r_j-\mathbf r_i|/\xi_e}.
\end{equation}
In polar coordinates, using rotational symmetry to fix the angular coordinate of
$\mathbf r_i$ and writing $\theta$ for the relative angle, this becomes
\begin{equation}
    \mathcal I(R)
    =
    2\pi
    \int_0^R r_i\,dr_i
    \int_R^\infty r_j\,dr_j
    \int_{-\pi}^{\pi} d\theta\,
    e^{-r_{ij}/\xi_e},
\end{equation}
with $  r_{ij}
    =
    \sqrt{
        r_i^2+r_j^2-2r_i r_j \cos\theta
    }$.
Equivalently,
\begin{equation}
    \boxed{
    \braket{W_A^\mathcal{Z}}
    \approx
    1
    -
    4\alpha^2
    \left[
    2\pi
    \int_0^R r_i\,dr_i
    \int_R^\infty r_j\,dr_j
    \int_{-\pi}^{\pi} d\theta\,
    e^{-r_{ij}/\xi_e}
    \right]
    } .
\end{equation}
The integral does not generally have a simple closed-form expression, but it can be
evaluated in two asymptotic limits.\\

{$\bullet$ Small-Loop Limit: $R\ll \xi_e$}\\

When $R\ll \xi_e$, the dominant outside endpoint satisfies $r_j \gg r_i$, so $r_{ij} \approx r_j$.
Then
\begin{align}
    \mathcal I(R)
    \approx
    \left(
        \int_{|\mathbf r_i|<R} d^2\mathbf r_i
    \right)
    \left(
        \int_{\mathbb R^2} d^2\mathbf r_j\,
        e^{-r_j/\xi_e}
    \right) =
    \pi R^2 \cdot 2\pi \xi_e^2 .
\end{align}
Therefore,
\begin{align}
    \braket{W_A^\mathcal{Z}}
    \approx
    1
    -
    4\alpha^2
    \left(
        \pi R^2
    \right)
    \left(
        2\pi \xi_e^2
    \right)
    =
    1 - 2n_e |A| .
\end{align}
This agrees with the area-law form
\begin{equation}
    \braket{W_A^\mathcal{Z}}
    \sim
    \exp\left(-2n_e |A|\right).
\end{equation}\\

{$\bullet$ Large-Loop Limit: $R \gg \xi_e$}\\

For $R \gg \xi_e$, the exponential decay implies that the dominant contribution
comes from strings whose endpoints lie close to the boundary of $A$. In other words, $ r_i, r_j \approx R$
Using the estimate $r_{ij}
    \le
    (r_j-r_i) + r_j\theta$,
we obtain the bound
\begin{align}
    \braket{W_A^\mathcal{Z}}
    &\ge
    1
    -
    4\alpha^2
    \int_0^R
    2\pi e^{r_i/\xi_e} r_i\,dr_i
    \int_R^\infty
    e^{-r_j/\xi_e} r_j\,dr_j
    \int_{-\pi}^{\pi}
    e^{-r_j\theta/\xi_e}\,d\theta .
\end{align}
Therefore,
\begin{align}
    \braket{W_A^\mathcal{Z}}
    \gtrsim
    1
    -
    4\alpha^2
    \left[
        2\pi
        \left(R\xi_e e^{R/\xi_e}\right)
        \left(R\xi_e e^{-R/\xi_e}\right)
        \left(\frac{2\xi_e}{R}\right)
    \right] =
    1
    -
    4\alpha^2
    \left(
        4\pi R\xi_e^3
    \right).
\end{align}
We now see that $\braket{W_A^\mathcal{Z}} \gtrsim 1 - 4n_e R\xi_e$. This lower bound is consistent with a perimeter-law decay,
\begin{equation}
    \braket{W_A^\mathcal{Z}}
    \gtrsim
    \exp\!\left(
        -C n_e \xi_e |\partial A|
    \right),
\end{equation}
where $C$ is a constant of order one.

From the two limits, the crossover from the area law at small loops to the perimeter law at larger ones happens at the loop size $R\sim \xi_e$. In Fig.~\ref{fig:extra_scaling}(d), we display the value of $\xi_e$ evaluated from the decay profile of open $\mathcal{X}$ strings, $\langle \mathcal{S}_L^{\mathcal{X}} \rangle \sim e^{-L/\xi_e}$, which in our numerics exceeds the lattice size (16 kagome links) at the end of the ramp $\delta/\Omega = 4.5$. This justifies the free $e$-gas picture we developed in the main text. 

\end{document}